\documentclass[10pt,aps,pra,twocolumn,nofootinbib]{revtex4-1}   
\usepackage{amsmath}    % need for subequations
\usepackage{amsfonts}  %note how statements can be commented out
\usepackage{amssymb}
\usepackage{graphicx}   % for figures
\usepackage{epstopdf}
\usepackage[dvips]{epsfig}

\usepackage{mathrsfs}
\usepackage[usenames,dvipsnames]{color}

\usepackage[colorlinks]{hyperref}

\usepackage{colortbl}

\definecolor{gray(x11gray)}{rgb}{0.75, 0.75, 0.75}

\definecolor{darksienna}{rgb}{0.24, 0.08, 0.08}

\definecolor{darkblue}{rgb}{0.0, 0.13, 0.35}

\newcommand{\ignore}[1]{}

\AtBeginDocument{%
    \hypersetup{
      urlcolor     = blue, %Color for external hyperlinks
      filecolor   = red,
      linkcolor    = blue, %Color of internal links
      citecolor   = blue, %Color of citations
    }
} % end  \AtBeginDocument

\begin{document}

%\title{Phase sensitivity and Fisher information in an unbalanced Mach-Zehnder interferometer fed with two inputs}

%\title{Phase sensitivity for an unbalanced interferometer without phase-matching restriction at its inputs}

\title{Phase sensitivity for an unbalanced interferometer without input phase-matching restrictions}

\author{Anca Preda}
\affiliation{Faculty of Physics, University of Bucharest, 077125 Bucharest-M\u{a}gurele, Romania}
%\email{ancaa.preda@yahoo.ro}
\email{anca.preda10@gmail.com}

\author{Stefan~Ataman}
\affiliation{Extreme Light Infrastructure - Nuclear Physics (ELI-NP), `Horia Hulubei' National R\&D Institute for Physics and Nuclear Engineering (IFIN-HH), 30 Reactorului Street, 077125 M\u{a}gurele, jud. Ilfov, Romania}
\email{stefan.ataman@eli-np.ro}

\date{\today}

% ----------------------------------------------------
% ----------------------- ABSTRACT ------------------------
% ----------------------------------------------------------

\begin{abstract}

The Cram\'er-Rao bound and the quantum Fisher information (QFI) have been tools used extensively for the interferometric phase sensitivity. Most scenarios considering a Mach-Zehnder interferometer (MZI) with two input sources focused on the phase-matched case, when the Fisher information is maximal. Under this constraint, the best sensitivity is achieved for a balanced (50/50) input beam splitter. In this paper, we take a different approach: we allow the beam splitter transmission coefficient as well as the input phase mis-match to be variable parameters. We then search for a pair of these parameters that maximizes the Fisher information. We find that for the double coherent input the maximum Fisher information can always be reached in the unbalanced case for a carefully chosen input phase mis-match. For the coherent plus squeezed vacuum case we find that under certain circumstances, a threshold phase mis-match exists, beyond which the optimum Fisher information is found for the degenerate case. For the squeezed-coherent plus squeezed vacuum case we find that the optimum is actually when the squeezing angles of the two inputs are in anti-phase.

\end{abstract}

\maketitle

% ---------------------------------------------------------
% ------------------- INTRODUCTION ------------------------
% ---------------------------------------------------------
\section{Introduction}
\label{sec:introduction}

Phase sensitivity is an old research topic that gained new momentum in recent years \cite{Par09,Gio11,Jar15,Ion15,LIGO13,Dem13,Sch17,Ata18b}, partly due to the emergence of quantum technologies \cite{Par09,Gio11,Jar15,Ion15} and of the gravitational wave astronomy \cite{LIGO13,Dem13,Sch17}. Among the long-established milestones one can mention the Cram\'er-Rao bound and the use of Fisher information \cite{Bra94} in order to obtain theoretical limits on parameter estimation \cite{Dem15}.

The well-known standard quantum limit (also called shot noise limit) was legion for an interferometric phase measurement until Caves' paper \cite{Cav81}, proving a way to go beyond it. The new sensitivity target using non-classical states of light called Heisenberg limit was investigated and proven fundamental \cite{Gio12}. Among the quantum states that reach this limit we mention the so-called NOON-states \cite{Hol93,Bot00,Cam03}, however losses can lead to a rapid degradation of this performance \cite{Dor09,Dem12}. Practical schemes focus mainly on squeezed states of light \cite{Yue76,Yur85,Xia87}. The advantage of these (squeezed) states lies in their ability to perform well in the low- as well as in the high-intensity regime \cite{LIGO13,Gar17,API18}.

While estimating the phase sensitivity of a Mach-Zehnder interferometer, practicalities arise. For example, the phase sensitivity is not uniform \cite{Dem15}. For a coherent input state, a workaround at low intensities has been shown to exist \cite{Pez07}. The actual detection scheme needs also to be taken into account, since different setups can lead to different sensitivities \cite{Dem15,Gar17,API18}.

However, when focusing on the Fisher information only, the detection scheme is disregarded \cite{Jar12,Tak17,Lan13,Lan14}. This is so because the Fisher information is always a best case scenario, hence its utility in finding the maximum theoretical performance one expects from a given setup.

Optimal phase sensitivity based on the quantum Fisher information is a subject amply discussed in the literature \cite{Dem15,API18,Jar12,Tak17,Liu13,Lan13,Lan14,Pez15}. This measure of information is not free of controversies or subtleties such as the influence of an external phase reference \cite{Jar12}, pathologies in the case of entangled input states \cite{Lan14} or no-go theorems correcting previous statements \cite{Tak17}.

It is generally believed that the balanced (50/50) beam splitter case is optimum and many works focus on this scenario only \cite{Hol93,Lan13,Lan14,Gar17}. Papers considering the non-balanced case reach this conclusion \cite{Jar12} or find no difference between the balanced and non balanced cases for the input phase matching condition \cite{Liu13}. 

We show in this paper that the phase-matched inputs coupled with a balanced interferometer is only a particular situation, from a much broader set of possibilities.

We discuss the double coherent input scenario \cite{API18,Shi99}, this time however with a non-balanced input beam splitter. Extending some previous results \cite{API18} for the balanced case, we find that there exists an optimum input phase mis-match different from zero for non-balanced interferometers. Moreover, we show that the maximum Fisher information stays the same, regardless of the input phase mis-match, if the transmission coefficient of the input beam splitter is carefully chosen.

For the often-discussed coherent plus squeezed vacuum input scenario, under certain circumstances, we find three different cases. For some given coherent amplitude and squeezing factor, we find a limit phase mis-match, $\Delta\theta_{\mathrm{lim}}$, that does not depend on the input beam splitter transmission factor. For phase mis-matches below this value, the Fisher information is indeed optimal in the balanced case. However, this assertion is no longer true beyond $\Delta\theta_{\mathrm{lim}}$.

The most general case of coherent squeezed $\otimes$ coherent squeezed (also called bright coherent) input was considered in the literature \cite{Spa15,Spa16}, however with some limitations. First, only the balanced case was considered. Second, a single input phase mis-match was allowed. Third, it used the single-parameter Fisher information approach, thereby counting resources that are actually not available. In our paper we discuss a slightly simplified scenario comprised of a squeezed-coherent source in one input and squeezed vacuum in the other. However, we impose neither a balanced beam splitter, nor any type of input phase matching. This scenario was discussed in the context of a balanced MZI and with difference-intensity detection in \cite{Par95}.

Throughout the paper we use the quantum Fisher information matrix (see references \cite{Lan13,Tak17} for a discussion) in order to avoid counting resources that are actually unavailable (see also \cite{Jar12}).

This paper is structured as follows. In Section \ref{sec:MZI_phase} we briefly introduced our experimental setup as well as the theoretical tools and conventions we use throughout this work. In Section \ref{sec:Fisher_double_coherent} we consider the double coherent case and discuss potential benefits from the non-balanced scenario. In Section \ref{sec:Fisher_coh_sqz_vacuum} we consider the coherent plus squeezed vacuum input and discuss the three possible cases one encounters. The more general case of squeezed-coherent plus squeezed vacuum input is detailed in Section \ref{sec:Fisher_szq_coh_sqz_vacuum}. The paper closes with the conclusions from Section \ref{sec:conclusions}.

% ---------------------------------------------------------
% --------------------- SECTION ---------------------------
% ---------------------------------------------------------
\section{Phase sensitivity with the Mach-Zehnder interferometer}
\label{sec:MZI_phase}

% ---------------------------------------------------------
% ---------------------------------------------------------
% ---------------------------------------------------------
\subsection{Field operator transformations}
\begin{figure}
\includegraphics[scale=0.75]{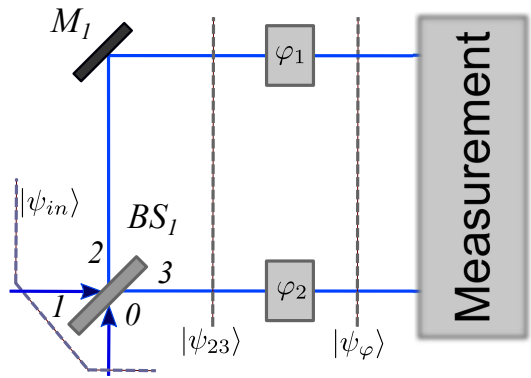}
\caption{The configuration for the case study with two independent phase shifts. The beam splitter $BS_1$ is assumed to have a variable transmission coefficient $T$. The wavevector used in all QFI calculations is given by equation~\eqref{eq:psi_phi_for_QFI}.}
\label{fig:MZi_Fisher_two_phases}
\end{figure}

We use a standard quantum optical description of our MZI \cite{GerryKnight} and we consider the operator transformations
\begin{equation}
\label{eq:field_op_transf_MZI_general}
\left\{
\begin{array}{l}
\hat{a}_3^\dagger=R^*\hat{a}_0^\dagger+T^*\hat{a}_1^\dagger\\
\hat{a}_2^\dagger=T^*\hat{a}_0^\dagger+R^*\hat{a}_1^\dagger
\end{array}
\right.
\end{equation}
where $T$ ($R$) denotes the transmission (reflection) coefficient of the beam splitter $BS_1$ (see Fig.~\ref{fig:MZi_Fisher_two_phases}). We have $\vert{T}\vert^2+\vert{R}\vert^2=1$ and $TR^*+T^*R=0$ \cite{GerryKnight} and the commutation relations $[\hat{a}_m,\hat{a}_n^\dagger]=\delta_{mn}$ with $m,n\in\{0,1\}$ and $\hat{a}_m$ ($\hat{a}_m^\dagger$) denotes the annihilation (creation) operator for the mode $m$. Equations \eqref{eq:field_op_transf_MZI_general} and the constraints stated before imply that the (photon) number operators ($\hat{n}_m=\hat{a}_m^\dagger\hat{a}_m$ for a mode $m$) are
\begin{equation}
\label{eq:a3_dagger_a3_versus_a1_a0}
\hat{n}_3=\vert{R}\vert^2\hat{a}_0^\dagger\hat{a}_0
+\vert{T}\vert^2\hat{a}_1^\dagger\hat{a}_1
-T^*R\left(\hat{a}_0^\dagger\hat{a}_1-\hat{a}_0\hat{a}_1^\dagger\right)
\end{equation}
and
\begin{equation}
\label{eq:a2_dagger_a2_versus_a1_a0}
\hat{n}_2=\vert{T}\vert^2\hat{a}_0^\dagger\hat{a}_0
+\vert{R}\vert^2\hat{a}_1^\dagger\hat{a}_1
+T^*R\left(\hat{a}_0^\dagger\hat{a}_1-\hat{a}_0\hat{a}_1^\dagger\right)
\end{equation}
Equations \eqref{eq:a3_dagger_a3_versus_a1_a0} and \eqref{eq:a2_dagger_a2_versus_a1_a0} allow to immediately find all needed output operator relations. For example by adding these equations we obtain $\hat{n}_3+\hat{n}_2
=\hat{a}_1^\dagger\hat{a}_1+\hat{a}_0^\dagger\hat{a}_0$, a statement that simply confirms the conservation of the number of photons.

When discussing the phase sensitivity of a MZI, one encounters several scenarios. In the single parameter scenario, we have a single phase shift, usually denoted by $\varphi$. Two sub-cases open now: we can assume this phase shift in a single arm of the interferometer, and using the notations from Fig.~\ref{fig:MZi_Fisher_two_phases} we have $\varphi_2=\varphi$, $\varphi_1=0$ and ${\vert\psi_\varphi\rangle=e^{-i\varphi_2\hat{n}_3}\vert\psi_{23}\rangle}$. Alternatively we can consider the phase shift split between the two arms. We have $\varphi_1=\varphi/2$ and $\varphi_2=-\varphi/2$, thus $\vert\psi_\varphi\rangle=e^{-i\frac{\varphi}{2}\hat{n}_3+i\frac{\varphi}{2}\hat{n}_2}\vert\psi_{23}\rangle$. In reference \cite{Jar12}, these cases were labelled (i) and (ii) and they corresponded to the Fisher informations $\mathcal{F}^{(i)}$ and $\mathcal{F}^{(ii)}$.

However, as discussed in the literature \cite{Jar12,Tak17,Lan13}, a two-parameter estimation technique avoids the problems of counting supplementary resources (like an external phase reference) that are actually not available.

The wavevector we consider, $\vert\psi_\varphi\rangle$, is expressed in respect with the sum/difference phase shifts ${\varphi_{s/d}=(\varphi_1\pm\varphi_2)/2}$, therefore
\begin{equation}
\label{eq:psi_phi_for_QFI}
\vert\psi_\varphi\rangle=e^{-i\frac{\hat{n}_2-\hat{n}_3}{2}\varphi_d}e^{-i\frac{\hat{n}_2+\hat{n}_3}{2}\varphi_s}\vert\psi_{23}\rangle
\end{equation}
and the state $\vert\psi_{23}\rangle$ is obtained by applying the field operator transformations \eqref{eq:field_op_transf_MZI_general} to the input state $\vert\psi_{in}\rangle$. 

Throughout this paper we consider that $\vert\psi_{in}\rangle$   is separable (\emph{i. e.} we assume no entanglement between the two inputs).

% ---------------------------------------------------------
% ---------------------------------------------------------
\subsection{The quantum Cram\'er-Rao bound and Fisher information}
\label{subsec:Fisher_matrix}
In parameter estimation a fundamental limit is given by the quantum Cram\'er-Rao bound (QCRB) \cite{Par09,Dem15,Lan13,Lan14}
\begin{equation}
(\Delta\varphi)^2\geq\frac{1}{\mathcal{F\left(\varphi\right)}}
\end{equation}
where $(\Delta\varphi)^2$ is the variance of the parameter $\varphi$ and $\mathcal{F\left(\varphi\right)}$ is the quantum Fisher information \cite{Dem15}.

In the two-parameter case the Fisher matrix \cite{Lan13} is given by
\begin{equation}
\label{eq:Fisher_Matrix}
\mathfrak{F}=\left(
\begin{array}{cc}
\mathfrak{F}_{ss} & \mathfrak{F}_{sd}\\
\mathfrak{F}_{ds} & \mathfrak{F}_{dd}
\end{array}
\right)
\end{equation}
where we define the matrix elements
\begin{equation}
\label{eq:Fisher_Matrix_elements_DEF}
\mathfrak{F}_{ij}=4\Re\left(\langle\partial_i\psi_\varphi\vert\partial_j\psi_\varphi\rangle
-\langle\partial_i\psi_\varphi\vert\psi_\varphi\rangle\langle\psi_\varphi\vert\partial_j\psi_\varphi\rangle\right)
\end{equation}
with $i,j=\{s,d\}$. The Cram\'er-Rao bound is now in matrix form ${\Sigma\geq\mathfrak{F}^{-1}}$, or, explicitly written
\begin{equation}
\label{eq:CRB_Fisher_Matrix}
\left(
\begin{array}{cc}
\Delta^2\varphi_{s} & \Delta\varphi_{s}\Delta\varphi_{d}\\
\Delta\varphi_{d}\Delta\varphi_{s} & \Delta^2\varphi_{d}
\end{array}
\right)
\geq\frac{1}{D_{\mathfrak{F}}}
\left(
\begin{array}{cc}
\mathfrak{F}_{dd} & -\mathfrak{F}_{sd}\\
-\mathfrak{F}_{ds} & \mathfrak{F}_{ss}
\end{array}
\right)
\end{equation}
where the determinant is $D_{\mathfrak{F}}=\mathfrak{F}_{ss}\mathfrak{F}_{dd}-\mathfrak{F}_{sd}\mathfrak{F}_{ds}$. In particular, for the phase difference sensitivity we have $\Delta\varphi_{d}\geq1/\sqrt{\mathcal{F}}$ where we defined
\begin{equation}
\label{eq:F_is_Fdd_minus_F_ds2_over_Fss}
\mathcal{F}=
\frac{\mathfrak{F}_{ss}\mathfrak{F}_{dd}-\mathfrak{F}_{sd}\mathfrak{F}_{ds}}{\mathfrak{F}_{ss}}
=\mathfrak{F}_{dd}-\frac{\mathfrak{F}_{sd}\mathfrak{F}_{ds}}{\mathfrak{F}_{ss}}
\end{equation}
For the remainder of the paper, when mentioning Fisher information we mean $\mathcal{F}$ and when mentioning phase sensitivity, we mean $\Delta\varphi_{d}$. For simplicity, we do not consider the effect of repeated measurements in this paper. In a nutshell, for $N$ identically prepared experiments we expect $\Delta\varphi_{d}\to\Delta\varphi_{d}/\sqrt{N}$ (see \cite{Pez07}).

Before ending this short section we remark that in equation \eqref{eq:F_is_Fdd_minus_F_ds2_over_Fss} we could have done the approximation ${\mathcal{F}\approx\mathfrak{F}_{dd}}$, as done in reference \cite{Lan13}. This would be well justified in the balanced case, since from most (although not all) input states one finds that $\mathfrak{F}_{sd}=\mathfrak{F}_{ds}=0$. In this paper we voluntarily avoid this restriction, thus the full expression from \eqref{eq:F_is_Fdd_minus_F_ds2_over_Fss} will be employed.

\begin{figure}
	\includegraphics[scale=0.45]{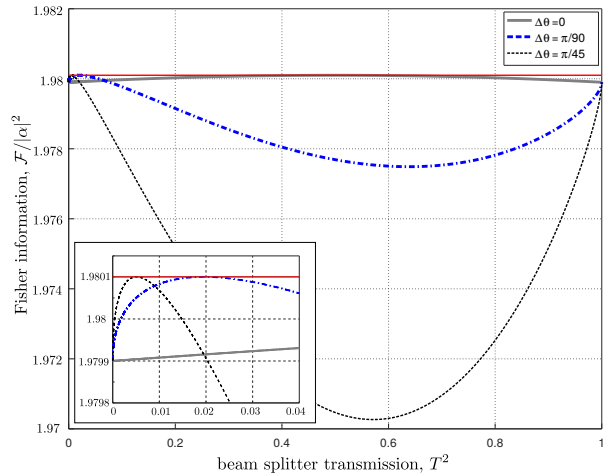}
	\caption{The Fisher information $\mathcal{F}$ from equation \eqref{eq:Fisher_dual_coherent} divided by $\vert\alpha\vert^2$ versus the beam splitter transmission coefficient ($T^2$) for three values of $\Delta\theta$. One notices that whatever the value of $\Delta\theta$, the Fisher information attains the same maximum $\mathcal{F}_{max}/\vert\alpha\vert^2=1+\varpi^2$, depicted with the solid red line. In the inset: zoom on the upper-left corner of the graphic. We used the values $\vert\alpha\vert=10$ and $\vert\beta\vert=9.9$.}
	\label{fig:Fisher_vs_T2_is constant_3theta}
\end{figure}

% -----------------------------------------------------
% --------------- SECTION -------- DOUBLE COHERENT --------
% ---------------------------------------------------------
\section{Fisher information for double coherent input}
\label{sec:Fisher_double_coherent}
The double coherent input scenario has been discussed at large in \cite{API18} for the balanced case. We extend them here to encompass also the non-balanced case. The input state is
\begin{equation}
\label{eq:psi_in_double_coherent}
\vert\psi_{in}\rangle=\vert\alpha_1\beta_0\rangle=\hat{D}_1\left(\alpha\right)\hat{D}_0\left(\beta\right)\vert0\rangle
\end{equation}
where the displacement operator \cite{GerryKnight} at input port $m$ is ${\hat{D}_m\left(\gamma\right)= e^{\gamma\hat{a}_m^\dagger-\gamma^*\hat{a}_m}}$ with $m=\{0,1\}$. Here $\alpha= \vert\alpha\vert e^{i\theta_\alpha}$, $\beta= \vert\beta\vert e^{i\theta_\beta}$ and $\Delta\theta= \theta_\alpha-\theta_\beta$ is the phase difference between the two input lasers.

We use the two-parameter approach to determine the Fisher information \cite{Lan13}. Calculations are detailed in Appendix \ref{sec:app:Fisher_double_coherent} and denoting $\varpi={\vert\beta\vert}/{\vert\alpha\vert}$ we have the final result
\begin{eqnarray}
\label{eq:Fisher_dual_coherent}
\mathcal{F}
=4\vert\alpha\vert^2\left(\vert{TR}\vert^2\left(1+\varpi^2\right)
-4\vert{TR}\vert^2\frac{\varpi^2\left(1+\sin^2\Delta\theta\right)}{1+\varpi^2}
\right.
% ------------- NEW LINE ---------
\nonumber\\
+\frac{\varpi^2}{1+\varpi^2}
%\nonumber\\
\left.
-2\vert{TR}\vert\left(\vert{T}\vert^2-\vert{R}\vert^2\right)\frac{\varpi\left(1-\varpi^2\right)\sin\Delta\theta}{1+\varpi^2}\right)
\quad
\end{eqnarray}
For a balanced beam splitter ($T=1/\sqrt{2}$ and $R=i/\sqrt{2}$) we arrive at
\begin{eqnarray}
\label{eq:Fisher_dual_coherent_BAL}
\mathcal{F}
=\vert\alpha\vert^2\left(1+\varpi^2
-\frac{4\varpi^2\sin^2\Delta\theta}{1+\varpi^2}
\right)
%\quad
\end{eqnarray}
a result reported in \cite{API18}. With this restriction, the optimum angle between the coherent input sources is obviously $\Delta\theta_{\mathrm{opt}}=k\pi$ with $k\in\mathbb{Z}$, which leads to maximum Fisher information,
\begin{eqnarray}
\label{eq:Fisher_dual_coherent_MAXIMUM}
{\mathcal{F}_{max}=\vert\alpha\vert^2(1+\varpi^2)}=\vert\alpha\vert^2+\vert\beta\vert^2
\end{eqnarray}
and to the phase sensitivity ${\Delta\varphi=1/\sqrt{\vert\alpha\vert^2(1+\varpi^2)}}$.

\begin{figure}
	\includegraphics[scale=0.45]{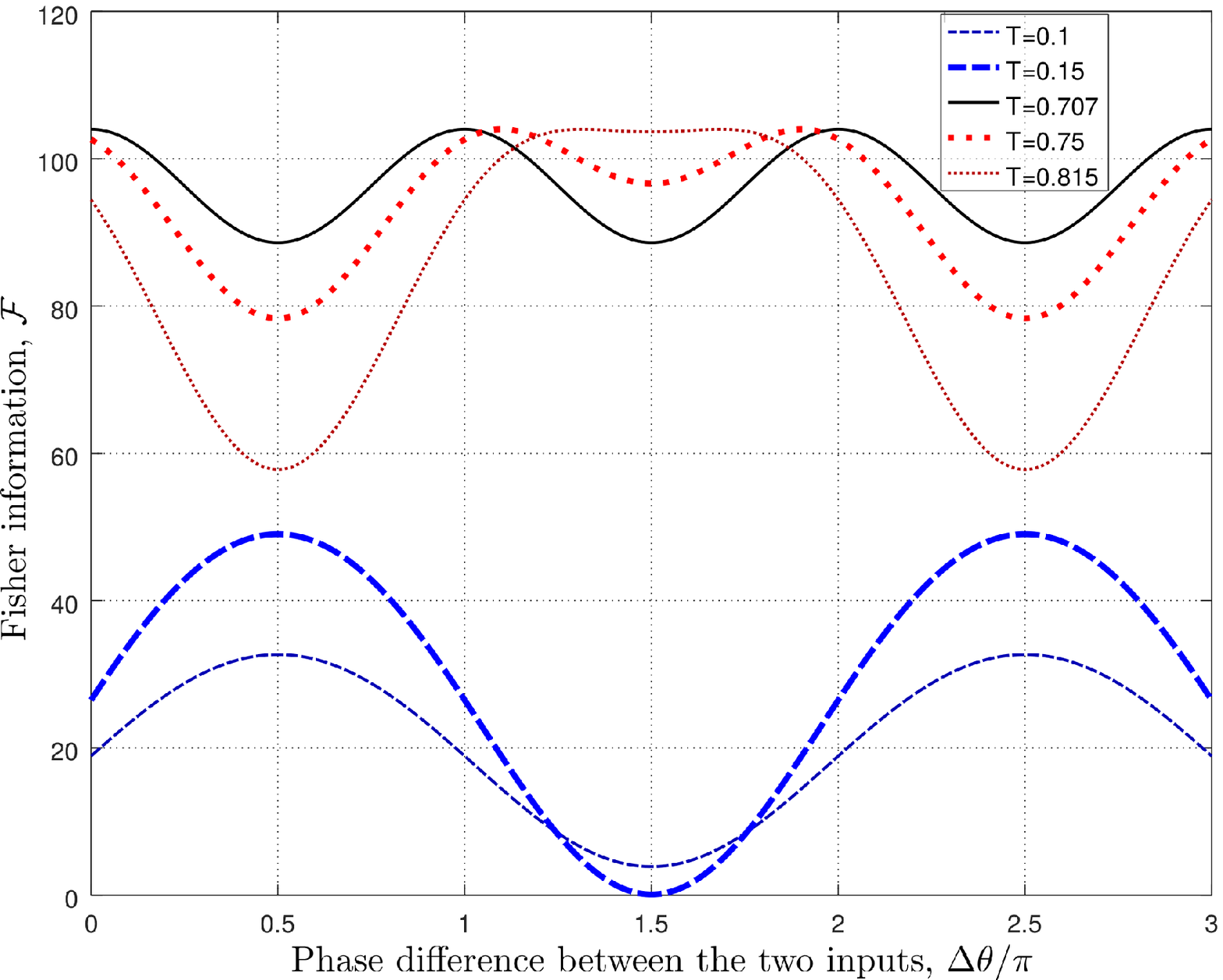}
	\caption{The Fisher information $\mathcal{F}$ versus the input phase mis-match, $\Delta\theta$, for five different values of $T$. We used the parameters: $\vert\alpha\vert=10$, $\vert\beta\vert=2$ (implying $\varpi=0.2$).}
	\label{fig:Fisher_dual_coh_versus_delta_theta_varpi02}
\end{figure}

We wish, however, to find the optimum input phase mis-match for the general, non-balanced case. Therefore, differentiating the expression from equation \eqref{eq:Fisher_dual_coherent} in respect with $\Delta\theta$ yields an optimum angle
\begin{equation}
\label{eq:Delta_theta_opt_dual_coh}
\Delta\theta_{\mathrm{opt}}=\arcsin\left(\frac{\vert{T}\vert^2-\vert{R}\vert^2}{4\vert{TR}\vert}\left(\frac{1}{\varpi}-1\right)\right)+2k\pi
\end{equation}
The significance of this (optimum) angle is the following: if we are constrained by a given transmission coefficient $T$, the maximum Fisher information $\mathcal{F}_{max}$ from equation \eqref{eq:Fisher_dual_coherent_MAXIMUM} can still be reached if the input phase mis-match is given by $\Delta\theta_{\mathrm{opt}}$ from equation \eqref{eq:Delta_theta_opt_dual_coh}. For the balanced case, one recovers $\Delta\theta_{\mathrm{opt}}=k\pi$, as expected.

The ability to reach the maximum sensitivity for a non-balanced beam splitter is an obvious practical advantage: small deviations in the transmission coefficient $T$ of a real-life beam splitter can be corrected via an easily adjustable phase shift $\Delta\theta$.

We try to answer now the following question: for a given input phase mis-match $\Delta\theta$ and for a fixed ratio $\varpi=\vert\beta\vert/\vert\alpha\vert$ what value of $T$ maximizes the Fisher information?

The calculations are detailed in Appendix \ref{app:sec:Calc_T_opt_versus_Delta_theta_dual_coh} and the final result is
\begin{equation}
\label{eq:T_opt_squared_versus_Delta_theta_dual_coh}
\vert{T}\vert^2_{\mathrm{opt}}=
\frac{1}{2}+
\frac{\mathrm{sign}(\varpi^2-1)\varpi \sin\Delta\theta}
{\sqrt{\left(1-\varpi^2\right)^2
+4\varpi^2\sin^2\Delta\theta}
}
\end{equation}
For simplicity, in the following plots we take $T$ real. In Fig.~\ref{fig:Fisher_vs_T2_is constant_3theta} we depict the Fisher information $\mathcal{F}$ from equation~\eqref{eq:Fisher_dual_coherent} versus the transmission coefficient ${T}^2$ for three different phase mis-matches $\Delta\theta$. It can be noted that in all three scenarios, the maximum Fisher information \eqref{eq:Fisher_dual_coherent_MAXIMUM} is reached at the optimal transmission coefficient ${T}^2_{\mathrm{opt}}$ given by equation \eqref{eq:T_opt_squared_versus_Delta_theta_dual_coh}. Thus, contrary to the balanced case from equation \eqref{eq:Fisher_dual_coherent_BAL}, when any nonzero input phase mis-match $\Delta\theta$ only degrades the performance, a tunable pair of parameters ($T$, $\Delta\theta$) can ensure that the best attainable phase sensitivity of the measurement is always reached.

\begin{figure}
	\includegraphics[scale=0.45]{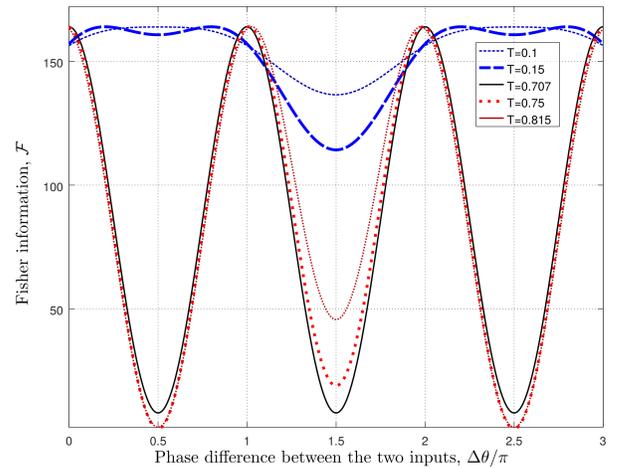}
	\caption{The Fisher information $\mathcal{F}$ versus the input phase mis-match, $\Delta\theta$, for five different values of $T$. We used the parameters: $\vert\alpha\vert=10$, $\vert\beta\vert=8$ (implying $\varpi=0.8$).}
	\label{fig:Fisher_dual_coh_versus_delta_theta_varpi08}
\end{figure}

We prepare to answer now a different question: if the input phase mis-match $\Delta\theta$ is fluctuating in certain intervals, can a well-chosen transmission coefficient $T$ compensate at some degree for this fluctuation? In order to answer this question, in Fig.~\ref{fig:Fisher_dual_coh_versus_delta_theta_varpi02} we plot the Fisher information versus the input phase difference $\Delta\theta$ for five different values of $T$. We took the scenario with $\varpi=0.2$ i. e. there is a big disparity between the powers of the input coherent sources. One notices that indeed, for the balanced case ($T=1/\sqrt{2}\approx0.707$) we have a maximum at $\Delta\theta=k\pi$ with $k\in\mathbb{Z}$. It is interesting to remark that for values $T>0.75$ the Fisher information remains almost constant for wide ranges of $\Delta\theta$ (for the case depicted in Fig.~\ref{fig:Fisher_dual_coh_versus_delta_theta_varpi02}, the interval is $\Delta\theta\in(\pi,2\pi)$). Thus, if an experimental setup is unable to keep a well determined phase shift between the two input sources [in the interval $\Delta\theta\in(\pi,2\pi)$], then choosing the right $T$ keeps the sensitivity almost constant.

The situation with almost equal intensity for the two input coherent sources ($\varpi=0.8$) is depicted in Fig.~\ref{fig:Fisher_dual_coh_versus_delta_theta_varpi08}. Once again, for the balanced case we find the optimum at $\Delta\theta=k\pi$. This time, however, the behaviour of the Fisher information radically changes. For $T>0.75$ we have a huge dependence of $\mathcal{F}$ on the angle $\Delta\theta$ while for small values of $T$ we have an almost constant Fisher information for a wide range of input phase shift differences [$\Delta\theta\in(0,\pi)$ for the scenario depicted in Fig.~\ref{fig:Fisher_dual_coh_versus_delta_theta_varpi08}].

\begin{figure}
	\includegraphics[scale=0.45]{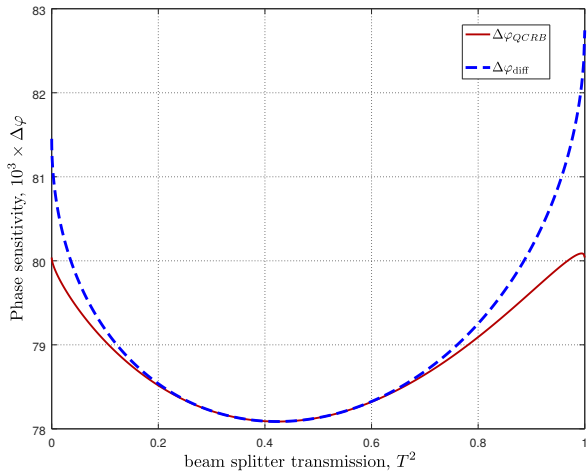}
	\caption{The phase sensitivity $\Delta\varphi$ for a dual coherent input versus the input beam splitter transmission factor $T$. We plotted the sensitivity from the quantum Cram\'er-Rao bound $\Delta\varphi_{QCRB}$ and the one from  the realistic difference-intensity detection scheme, $\Delta\varphi_{\mathrm{diff}}$. We used the parameters: $\vert\alpha\vert=10$, $\vert\beta\vert=8$, $\Delta\theta=\pi/90$. We optimized $\varphi_\mathrm{opt}$ from equation \eqref{eq:app:theta_opt_dial_coh_nonbal} for a value $\vert{T}\vert^2=0.25$.}
	\label{fig:Delta_phi_dual_coh_QCRB_diff_det}
\end{figure}

Before concluding this section we would like to investigate if the theoretically computed sensitivities are achievable in a realistic detection scheme. We therefore choose the difference intensity detection scheme (see e. g. Fig.~2 in \citep{API18} and Appendix \ref{sec:app:dual_coh_diff_det}).  Using standard techniques, we find the phase sensitivity
%\begin{widetext}
\begin{eqnarray}
\label{eq:Delta_phi_dual_coh_diff_det}
\Delta\varphi_{\mathrm{diff}}
=\frac{\sqrt{1+\varpi^2}}
{2\vert\alpha\vert\cdot\vert C_d\sin\varphi
+\varpi\cos\Delta\theta\cos\varphi
\vert}
\end{eqnarray}
%\end{widetext}
where $\varphi$ is the total internal phase in the interferometer and the coefficient $C_d$ is defined in equation \eqref{eq:app:Notation_C_d}.

For a given $T$ and $\Delta\theta$, there exists an optimum phase shift $\varphi_{opt}$ (see Appendix \ref{sec:app:dual_coh_diff_det}) so that $\Delta\varphi_{\mathrm{diff}}$ attains the quantum Cram\'er-Rao bound $\Delta\varphi_{QCRB}=1/\sqrt{\mathcal{F}}$ where $\mathcal{F}$ is given by equation \eqref{eq:Fisher_dual_coherent}.

We compare the best achievable sensitivity $\Delta\varphi_{QCRB}$ with the realistic one given by equation \eqref{eq:Delta_phi_dual_coh_diff_det} in Fig.~\ref{fig:Delta_phi_dual_coh_QCRB_diff_det}. We optimized our total internal phase $\varphi_\mathrm{opt}$ from equation \eqref{eq:app:theta_opt_dial_coh_nonbal} for $\vert{T}\vert^2=0.25$. Indeed, one can notice that in the vicinity of the optimized value we have an overlap between the realistic performance and the theoretically predicted one. Even for values of $T$ far from the optimized one, the performance does not degrade noticeably. We conclude that the theoretically predicted performances are actually achievable in realistic experiments.

% ---------------------------------------------------------
% ---------- SECTION --- COHERENT plus SQUEEZED VACUUM ----
% ---------------------------------------------------------
\section{Fisher information for coherent plus squeezed vacuum input}
\label{sec:Fisher_coh_sqz_vacuum}
In this rather popular scenario \cite{Cav81,Lan13, Xia87,Dem13} we consider the input state
\begin{equation}
\label{eq:psi_in_coherent_squeezed}
\vert\psi_{in}\rangle=\vert\alpha_1\xi_0\rangle=\hat{D}_1\left(\alpha\right)\hat{S}_0\left(\xi\right)\vert0\rangle
\end{equation}
The squeezed vacuum state acting on a port $m$ can be mathematically described by applying the squeezing operator ${\hat{S}_m\left(\xi\right)=e^{[\xi^*\hat{a}_m^2-\xi(\hat{a}_m^\dagger)^2]/2}}$ with ${\xi=re^{i\theta}}$. The parameter $r\in\mathbb{R}^+$ is usually called the squeezing factor. We define the input phase mis-match by $\Delta\theta=2\theta_\alpha-\theta$.

For a two-parameter Fisher information we have the matrix elements $\mathfrak{F}_{ij}$ with $i,j\in\{s,d\}$ given in  Appendix \ref{sec:app:Fisher_calcualtion_ocherent_squeezed_vacuum}). The final expression of the Fisher information is
\begin{eqnarray}
\label{eq:Fisher_coh_squeezed_vacuum_general}
\mathcal{F}=4\vert{TR}\vert^2\left(\vert\alpha\vert^2\left(\sinh2r\cos\Delta\theta+\cosh2r\right)+\sinh^2r
\right.
\nonumber\\
\left.
-\frac{2\sinh^22r\vert\alpha\vert^2}
{\vert\alpha\vert^2+\frac{\sinh^22r}{2}}\right)
+\frac{2\sinh^22r\vert\alpha\vert^2}
{\vert\alpha\vert^2+\frac{\sinh^22r}{2}}
\quad
\end{eqnarray}
For the balanced case we obtain
\begin{eqnarray}
\label{eq:Fisher_coh_squeezed_vacuum_balanced}
\mathcal{F}=\vert\alpha\vert^2\left(\sinh2r\cos\Delta\theta+\cosh2r\right)+\sinh^2r
\end{eqnarray}
giving the maximum achievable Fisher information
\begin{eqnarray}
\label{eq:Fisher_coh_squeezed_vacuum_balanced_MAX}
\mathcal{F}_{max}=\vert\alpha\vert^2e^{2r}+\sinh^2r
\end{eqnarray}
when the input phase matching condition is 
\begin{equation}
\label{eq:phase_matching_condition_coh_plus_sqz_vac}
\Delta\theta=2\theta_\alpha-\theta=0
\end{equation}
a conclusion reached by other papers, too \cite{Gar17,API18,Liu13}. Accordingly, we obtain the phase sensitivity ${\Delta\varphi=1/\sqrt{\vert\alpha\vert^2e^{2r}+\sinh^2r}}$, a result amply discussed in the literature \cite{Lan13,Jar12,Pez08}.

\begin{figure}
	\includegraphics[scale=0.45]{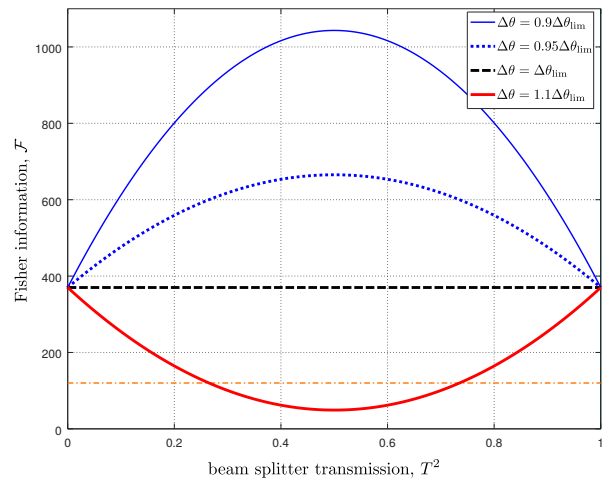}
	\label{fig:Fisher_vs_T2_coh_sqz_vac}
	\caption{The Fisher information versus the beam splitter transmission factor ($T^2$). For phase mis-match angles $\Delta\theta<\Delta\theta_{\mathrm{lim}}$ the optimum is found for a balanced beam splitter. For $\Delta\theta=\Delta\theta_{\mathrm{lim}}$ the Fisher information is constant regarless of the value of $T$ and given by equation \eqref{eq:Fisher_coh_squeezed_vacuum_degenerate}. For input angles $\Delta\theta$ beyond $\Delta\theta_{\mathrm{lim}}$, the optimum Fisher information is found for the degenerate case $T=0/1$. We used the values $\vert\alpha\vert=10$ and $r=2.3$ so that $\Delta\theta_{\mathrm{lim}}\approx0.9\pi$. For reference, we also plotted (orange, dash-dotted) the Fisher information for a single coherent input having the same average number of photons (i. e. $\vert\alpha'\vert^2=\vert\alpha\vert^2+\sinh^2r$).}
\end{figure}

At first glance, the Fisher information from equation \eqref{eq:Fisher_coh_squeezed_vacuum_general} is always maximized in the balanced case since it is well known that $\vert{TR}\vert$ is maximal in this case. However, the coefficient multiplying this factor is not necessarily positive for any combination of $\alpha$, $r$ and $\theta$. Thus, denoting by $\kappa$ the term multiplying $\vert{TR}\vert^2$ in equation \eqref{eq:Fisher_coh_squeezed_vacuum_general} i. e.
\begin{eqnarray}
\kappa=\vert\alpha\vert^2\left(\sinh2r\cos\Delta\theta+\cosh2r\right)+\sinh^2r
\nonumber\\
-\frac{2\sinh^22r\vert\alpha\vert^2}
{\vert\alpha\vert^2+\frac{\sinh^22r}{2}}
\end{eqnarray}
three cases can be distinguished: (i) $\kappa>0$, (ii) $\kappa=0$ and (iii) $\kappa<0$.

In the case (i) we obviously have a best case scenario for a balanced beam splitter yielding the expression from equation \eqref{eq:Fisher_coh_squeezed_vacuum_balanced}. Moreover, the Fisher information is further maximized for $\Delta\theta=2k\pi$ with $k\in\mathbb{Z}$ and we arrive at the known result from equation \eqref{eq:Fisher_coh_squeezed_vacuum_balanced_MAX}.

In the case (ii), the Fisher information reduces to
\begin{eqnarray}
\label{eq:Fisher_coh_squeezed_vacuum_degenerate}
\mathcal{F}=\frac{2\sinh^22r\vert\alpha\vert^2}
{\vert\alpha\vert^2+\frac{\sinh^22r}{2}}
\end{eqnarray}
irrespective of the value of $T$. This condition is equivalent to having an input phase mis-match $\Delta\theta_{\mathrm{lim}}$ that obeys the equation
\begin{equation}
\label{eq:cos_Delta_theta_lim_DEF}
\cos\Delta\theta_{\mathrm{lim}}=
\frac{2\sinh2r}
{\vert\alpha\vert^2+\frac{\sinh2r}{2}}
-\frac{\cosh2r}{\sinh2r}-\frac{\sinh^2r}{\sinh2r\vert\alpha\vert^2}
\end{equation}
Solutions $\Delta\theta_{\mathrm{lim}}$ exist as long as the r.h.s. of equation \eqref{eq:cos_Delta_theta_lim_DEF} is between $-1$ and $1$. Thus, not all values of $\vert\alpha\vert$ and $r$ can lead to the Fisher information from equation \eqref{eq:Fisher_coh_squeezed_vacuum_degenerate}. Notheworthy, $\Delta\theta_{\mathrm{lim}}$ is a function of only $\vert\alpha\vert$ and $r$, the value of $T$ playing no role. For the experimentally interesting scenario $\vert\alpha\vert^2\gg\sinh^2r$ we can approximate equation \eqref{eq:cos_Delta_theta_lim_DEF} to
\begin{equation}
\Delta\theta_{\mathrm{lim}}\approx
\arccos\left(\frac{2\sinh2r}
{\vert\alpha\vert^2}
-\frac{\cosh2r}{\sinh2r}\right)
\end{equation}
where we assume that the argument of the inverse cosine is between $-1$ and $1$.

In the case (iii) we can write the Fisher information as
\begin{eqnarray}
\label{eq:Fisher_coh_squeezed_vacuum_degenerate_negative}
\mathcal{F}=-4\vert{TR}\vert^2\vert\kappa\vert+\frac{2\sinh^22r\vert\alpha\vert^2}
{\vert\alpha\vert^2+\frac{\sinh^22r}{2}}
\end{eqnarray}
and now the maximum value of $\mathcal{F}$ [given by equation \eqref{eq:Fisher_coh_squeezed_vacuum_degenerate}] is reached when $\vert{TR}\vert^2$ is minimal, implying the degenerate case $T=0/1$.

In Fig.~\ref{fig:Fisher_vs_T2_coh_sqz_vac} we plot the Fisher information for the three cases discussed before. For input phase mis-matches $\Delta\theta<\Delta\theta_{\mathrm{lim}}$ the Fisher information reaches its maximum for $T^2=1/{2}$, as expected. However, as $\Delta\theta\to\Delta\theta_{\mathrm{lim}}$, the Fisher information approaches the value given by equation \eqref{eq:Fisher_coh_squeezed_vacuum_degenerate} and beyond it, we find ourselves in the degenerate case, where the optimum is given when $T=0/1$.

Therefore, from an experimental point of view, in order to maximize the sensitivity, it is important to keep the input phase mis-match as small as possible, ideally ${\Delta\theta=0}$. If this is not possible, a degradation of the sensitivity is to be expected, with a possible threshold value at $\Delta\theta_{\mathrm{lim}}$ (depending on the values of $\vert\alpha\vert$ and $r$).
% Beyond this value, the interferometer is unusable.

% -----------------------------------------------------
% ------ SECTION --- COHERENT plus SQUEEZED VACUUM --------
% -------------------------------------------------------
\section{Fisher information for squeezed-coherent plus squeezed vacuum input}
\label{sec:Fisher_szq_coh_sqz_vacuum}
The most general scenario with Gaussian input states is obtained by applying squeezed-coherent states in both inputs i. e. $\vert\psi_{in}\rangle=\vert{(\beta\xi)_0}{(\alpha\zeta)_1}\rangle$, ${\xi=re^{i\theta}}$ and ${\zeta=ze^{i\phi}}$ where $r,z\in\mathbb{R}^+$ and $\theta,\phi\in[0,2\pi]$.

In this section we focus on a slightly simpler version of this state by setting $\beta=0$ i. e. in input $0$ we apply squeezed vacuum only,
\begin{equation}
\label{eq:psi_in_coh_squeezed_times_squeezed_vac}
\vert\psi_{in}\rangle
={\hat{S}_0\left(\xi\right)}{\hat{D}_1\left(\alpha\right)}{\hat{S}_1\left(\zeta\right)}\vert0\rangle
\end{equation}
The elements for the Fisher matrix are detailed in Appendix \ref{sec:app:Fisher_calcualtion_coh_sqz_plus_sqz_vac} and the final expression of the Fisher information is given by equation \eqref{eq:F_final_coh_sqz_vac_sqz_vac}. The input phase matching condition that maximizes the Fisher information is given by
\begin{equation}
\label{eq:Phase_matching_cond_sqz_coh_plus_sqz_vac}
\left\{
\begin{array}{l}
2\theta_\alpha-\theta=0\\
2\theta_\alpha-\phi=\pi+2k\pi\\
{\phi}-{\theta}=\pi+2k\pi
\end{array}
\right.
\end{equation}
with $k\in\mathbb{Z}$. Similar to the discussion from the previous section, we define  $\kappa$ given by equation \eqref{eq:app:kappa_zero_coh_sqz_vac_sqz_vac}. For $\kappa>0$ the optimum is found in the balanced case and we have
\begin{eqnarray}
\label{eq:F_final_coh_sqz_vac_sqz_vac_BALANCED}
\mathcal{F}
=\vert\alpha\vert^2\left(\cosh{2r}
+\sinh{2r}\cos\Delta\theta
\right)
+\sinh^2{r}+\sinh^2{z}
\quad
\nonumber\\
% ---------------------------
+2\sinh{r}\sinh{z}
\left(\sinh{r}\sinh{z}
-\cosh{r}\cosh{z}\cos(\phi-\theta)
\right)
\qquad
\end{eqnarray}
Compared to the Fisher information from the previous scenario \eqref{eq:Fisher_coh_squeezed_vacuum_balanced}, the increase is potentially larger, due to the last term of equation \eqref{eq:F_final_coh_sqz_vac_sqz_vac_BALANCED}. If we also reinforce the input phase matching conditions from equation \eqref{eq:Phase_matching_cond_sqz_coh_plus_sqz_vac} (noteworthy, the two squeezing angles have to be in anti-phase), the optimum Fisher information from equation \eqref{eq:F_final_coh_sqz_vac_sqz_vac_BALANCED} yields
\begin{equation}
\label{eq:F_MAX_sqz_coh_plus_squeezed_vac}
\mathcal{F}_{max}
%=\vert\alpha\vert^2e^{2r}+\frac{\cosh2(r+z)-1}{2}
=\vert\alpha\vert^2e^{2r}+\sinh^2(r+z)
\end{equation}

\begin{figure}
	\includegraphics[scale=0.45]{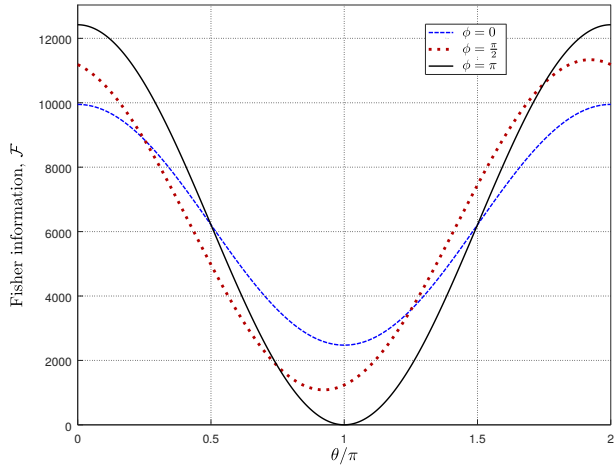}
	\caption{The Fisher information $\mathcal{F}$ from equation \eqref{eq:F_final_coh_sqz_vac_sqz_vac_BALANCED} versus the angle $\theta$ of the squeezed vacuum from input $0$ for three values of the phase $\phi$. We used the parameters: ${\vert\alpha\vert=10}$, $r=2.3$, $z=2.3$ and $\theta_\alpha=0$.}
	\label{fig:F_sqz_coh_sqz_vac_versus_phi_2theta_bal}
\end{figure}

In Fig.~\ref{fig:F_sqz_coh_sqz_vac_versus_phi_2theta_bal} we plot the Fisher information from equation \eqref{eq:F_final_coh_sqz_vac_sqz_vac_BALANCED} versus the angle $\theta$ for three difference values of $\phi$. For simplicity, we set $\theta_\alpha=0$. One notes that, indeed, the maximum is reached for $\theta=2k\pi$ with $k\in\mathbb{Z}$ and $\phi=\pi$. Thus, the reflex of considering the optimum scenario to be always the one with no input phase mis-match should be questioned for each new setup.

Similar to the discussion from Section \ref{sec:Fisher_coh_sqz_vacuum}, we can label with (ii) the case when $\kappa=0$ yielding a Fisher information independent on the value of the transmission coefficient $T$,
\begin{eqnarray}
\label{eq:F_final_coh_sqz_vac_sqz_vac_INDEP_T}
\mathcal{F}
=2\frac{\sinh^2{2r}
\left(
\frac{\sinh^2{2z}}{2}
+\vert\alpha\vert^2\left(\cosh{2z}+\sinh{2z}\cos{\Delta\phi}\right)\right)}
{\frac{\sinh^2{2r}}{2}
+\frac{\sinh^2{2z}}{2}+\vert\alpha\vert^2\left(\cosh{2z}+\sinh{2z}\cos\Delta\phi\right)}.
\nonumber\\
\mbox{ }
%\:\quad
\end{eqnarray}
From the condition $\kappa=0$ one can obtain a threshold angle $\Delta\theta_{\mathrm{lim}}$, this time however this value depends on $\alpha$, $r$, $z$ and $\phi$. Thus, the condition $\kappa=0$ can be satisfied by a myriad of scenarios implying all the aforementioned parameters.

For $\kappa<0$ find ourselves again in the degenerate case, when the optimum Fisher information is obtained when $T=0/1$.

Before ending our discussion, we make a brief comparison among the three scenarios we discussed while considering the values used throughout the paper: $\vert\alpha\vert=10$, $r=2.3$ and $z=2.3$. For the coherent plus squeezed vacuum scenario discussed in Section \ref{sec:Fisher_coh_sqz_vacuum} the maximum Fisher information is given by equation  \eqref{eq:Fisher_coh_squeezed_vacuum_balanced_MAX} and we have the value ${\mathcal{F}_{max}\approx9972}$ with an average number of photons $\langle{n}\rangle=\vert\alpha\vert^2+\sinh^2r\approx124$. If we assume the same average number of photons for the double coherent scenario discussed in Section \ref{sec:Fisher_double_coherent} we end up, according to equation \eqref{eq:Fisher_dual_coherent_MAXIMUM}, with ${\mathcal{F}_{max}\approx124}$. Now adding squeezing to input $1$, as discussed in this section brings us to ${\mathcal{F}_{max}\approx12422}$ with an average number of photons ${\langle{n}\rangle=\vert\alpha\vert^2+\sinh^2r+\sinh^2z\approx149}$. The advantage of adding squeezing is obvious, even for moderate squeezing factors.

% ---------------------------------------------------------
% --------------------- SECTION ---------------------------
% ---------------------------------------------------------
\section{Conclusion}
\label{sec:conclusions}
In this paper we considered the general case of an unbalanced MZI with three input scenarios: the double coherent, the coherent plus squeezed vacuum and the squeezed-coherent plus squeezed vacuum cases.

For the double coherent input we showed that the maximum Fisher information $\mathcal{F}_{max}=\vert\alpha\vert^2+\vert\beta\vert^2$ can always be reached, even if $\Delta\theta\neq0$, with a suitably unbalanced beam splitter. We showed that the availability of two new parameters (the input phase mis-match, $\Delta\theta$ and the beam splitter transmission coefficient, $T$) brings extra degrees of freedom for various experimental scenarios.

For the coherent plus squeezed vacuum case we find that in general the optimum is reached for the balanced case and with no input phase mis-match i .e. $\Delta\theta=0$. If one cannot meet the requirement $\Delta\theta=0$, under certain circumstances, there exists a threshold input phase mis-match, $\Delta\theta_{\mathrm{lim}}$. At this threshold, the Fisher information remains constant, irrespective of the beam splitter transmission coefficient. Beyond this threshold value, the optimum Fisher information is found for the degenerate case ($T=0/1$).

In the more general squeezed-coherent plus squeezed vacuum input case we find that the maximum Fisher information is in the balanced case with $\Delta\theta=0$ and with $\theta-\phi=\pi$ i. e. the two squeezing angles must be in anti-phase.

To conclude, when faced with a given input scenario, one must question when the optimum Fisher information is actually achieved and not disregard input phase mis-matches or unbalanced interferometers. 

\begin{acknowledgments}

S.A. acknowledges that this work has been supported by the Extreme Light Infrastructure Nuclear Physics (ELI-NP) Phase II, a project co-financed by the Romanian Government and the European Union through the European Regional Development Fund and the Competitiveness Operational Programme (1/07.07.2016, COP, ID 1334).

\end{acknowledgments}

\appendix

% ---------------------------------------------------------
% ---------------------------------------------------------
\section{Fisher information computations for the double coherent input}
\label{sec:app:Fisher_double_coherent}
In order to compute the Fisher information for the double coherent input, we employ a two-parameter approach described in Subsection \ref{subsec:Fisher_matrix}. For this computation, we need all four elements of the Fisher matrix from equation~\eqref{eq:Fisher_Matrix}. 
The sum-sum element is defined as $\mathfrak{F}_{ss}=4\Re\left(\langle\partial_s\psi\vert\partial_s\psi\rangle
-\vert\langle\psi\vert\partial_s\psi\rangle\vert^2\right)$.
Using the field operator transformations from equation~\eqref{eq:field_op_transf_MZI_general}, we have
\begin{equation}
\label{eq:Fisher_info_del_psi_s_del_phi_psi_s_two_param_DCoh}
\langle\partial_s\psi\vert\partial_s\psi\rangle
%=\frac{1}{4}\langle\psi_{in}\vert\left({\blue\hat{a}_1^\dagger\hat{a}_1^\dagger\hat{a}_1\hat{a}_1}+{\blue\hat{a}_1^\dagger\hat{a}_1}+2{\blue\hat{a}_1^\dagger\hat{a}_1}{\red\hat{a}_0^\dagger\hat{a}_0}+{\red\hat{a}_0^\dagger\hat{a}_0^\dagger\hat{a}_0\hat{a}_0}+{\red\hat{a}_0^\dagger\hat{a}_0}\right)\vert\psi_{in}\rangle
=\vert\alpha\vert^4+\vert\alpha\vert^2+2\vert\alpha\vert^2\vert\beta\vert^2+\vert\beta\vert^4+\vert\beta\vert^2
\end{equation}
The second term of $\mathfrak{F}_{ss}$ is 
\begin{equation}
\langle\psi\vert\partial_s\psi\rangle
=-{i}/{2}\left(\vert\alpha\vert^2+\vert\beta\vert^2\right),
\end{equation}
therefore the Fisher matrix element is $\mathfrak{F}_{ss}=\vert\alpha\vert^2+\vert\beta\vert^2$.

The difference-difference Fisher matrix element $\mathfrak{F}_{dd}$ is defined as
$\mathfrak{F}_{dd}=4\Re\left(\langle\partial_d\psi\vert\partial_d\psi\rangle
-\vert\langle\psi\vert\partial_d\psi\rangle\vert^2\right)$.
The first term is found to be
\begin{eqnarray}
\langle\partial_d\psi\vert\partial_d\psi\rangle=\vert{R\alpha+T\beta}\vert^4+\vert{R\alpha+T\beta}\vert^2
+\vert{T\alpha+R\beta}\vert^4 
\nonumber\\
+\vert{T\alpha+R\beta}\vert^2-2\vert{T\alpha+R\beta}\vert^2\vert{R\alpha+T\beta}\vert^2
\end{eqnarray}
while the second one gives
\begin{eqnarray}
\langle\psi\vert\partial_{d}\psi\rangle=\vert{R\alpha+T\beta}\vert^2-\vert{T\alpha+R\beta}\vert^2.
\end{eqnarray}
We thus have the Fisher matrix element $\mathfrak{F}_{dd}=\vert\alpha\vert^2+\vert\beta\vert^2$. 

The two remaining terms from the Fisher matrix are equal \cite{Lan13},
\begin{equation}
\mathfrak{F}_{sd}=\mathfrak{F}_{ds}
=\vert{R\alpha+T\beta}\vert^2-\vert{T\alpha+R\beta}\vert^2
%=\left(\vert{R}\vert^2-\vert{T}\vert^2\right)\left(\vert\alpha\vert^2-\vert\beta\vert^2\right)+4iTR^*\vert\alpha\beta\vert\sin\left(\theta_\beta-\theta_\alpha\right)
\end{equation}
and we can construct now the Fisher matrix,
\begin{widetext}
\begin{equation}
\label{eq:Fisher_Matrix_dual_coherent}
\mathfrak{F}=\left(
\begin{array}{cc}
\vert\alpha\vert^2+\vert\beta\vert^2 & \vert{R\alpha+T\beta}\vert^2-\vert{T\alpha+R\beta}\vert^2\\
\vert{R\alpha+T\beta}\vert^2-\vert{T\alpha+R\beta}\vert^2 & \vert\alpha\vert^2+\vert\beta\vert^2
\end{array}
\right)
\end{equation}
Its determinant is $D_{\mathfrak{F}}=4\vert{R\alpha+T\beta}\vert^2\vert{T\alpha+R\beta}\vert^2$. We define the Fisher information corresponding to a difference-difference measurement as $\mathcal{F}=D_{\mathfrak{F}}/\mathfrak{F}_{ss}$. After some calculations we have
\begin{eqnarray}
\label{eq:Fisher_dual_coh_alpha_beta}
\mathcal{F}
=4\vert{TR}\vert^2\left(\vert\alpha\vert^2+\vert\beta\vert^2\right)
-16\vert{TR}\vert^2\frac{\vert\alpha\beta\vert^2\sin^2\Delta\theta}{\vert\alpha\vert^2+\vert\beta\vert^2}
+4\left(\vert{T}\vert^2-\vert{R}\vert^2\right)^2\frac{\vert\alpha\beta\vert^2}{\vert\alpha\vert^2
+\vert\beta\vert^2}
\nonumber\\
% ------------- NEW LINE ---------
-8\vert{TR}\vert\left(\vert{T}\vert^2-\vert{R}\vert^2\right)\vert\alpha\beta\vert\frac{\vert\alpha\vert^2-\vert\beta\vert^2}{\vert\alpha\vert^2+\vert\beta\vert^2}\sin\Delta\theta
\quad
\end{eqnarray}
where we used the fact that $( \pm iT^*R)^2=\vert{TR}\vert^2$ and in this paper we made the convention $iT^*R=-\vert{TR}\vert$.

% --------------------------------------------------------
% ------------ DUAL COHERENT ------- coefficient T---------
% --------------------------------------------------------
\section{Calculation of the optimum transmission coefficient for the double coherent scenario}
\label{app:sec:Calc_T_opt_versus_Delta_theta_dual_coh}

We compute the transmission $\vert{T}\vert^2$ in terms of $\Delta\theta$, for which the Fisher information reaches its maximal value of $\mathcal{F}_{max}=\vert\alpha\vert^2(1+\varpi^2)$. We start from the general expression of the Fisher information given in equation \eqref{eq:Fisher_dual_coherent} and denote $x=\vert{TR}\vert$. Using the identity $\vert{T}\vert^2-\vert{R}\vert^2=\pm
\sqrt{1-4\vert{TR}\vert^2}$, we get
\begin{eqnarray}
\label{eq:Fisher_dual_coh_varpi}
\mathcal{F}
=4\vert\alpha\vert^2\left(x^2\frac{\left(1+\varpi^2\right)^2
-4\varpi^2\left(1+\sin^2\Delta\theta\right)}{1+\varpi^2}
+\frac{\varpi^2}{1+\varpi^2}
%\nonumber\\
\mp2x\sqrt{1-4x^2}\frac{\varpi\left(1-\varpi^2\right)\sin\Delta\theta}{1+\varpi^2}\right)
\end{eqnarray}
We define the coefficients:
$\alpha_x=4\vert\alpha\vert^2\frac{\left(1+\varpi^2\right)^2
-4\varpi^2\left(1+\sin^2\Delta\theta\right)}{1+\varpi^2}$, $\beta_x=4\vert\alpha\vert^2\frac{\varpi^2}{1+\varpi^2}$ and ${\delta_x=\mp8\vert\alpha\vert^2\frac{\varpi\left(1-\varpi^2\right)\sin\Delta\theta}{1+\varpi^2}}$, yielding 
\begin{eqnarray}
\label{eq:app:Fisher_2coh_second_degree_x}
\mathcal{F}=\alpha_xx^2
+\beta_x+x\sqrt{1-4x^2}\delta_x
\end{eqnarray}
We differentiate equation \eqref{eq:app:Fisher_2coh_second_degree_x} with respect to $x$ and after some computations we arrive at
%\begin{eqnarray}
$4(4\delta_x^2+\alpha_x^2)x^4
-(4\delta_x^2+\alpha_x^2)x^2
+{\delta_x^2}/{4}=0$.
%\end{eqnarray}
The solutions of this second degree equation are
\begin{eqnarray}
\label{eq:X^2_solutions}
x_{1,2}^2=\frac{1}{8}\left(1\pm\frac{1}{\sqrt{4\frac{\delta_x^2}{\alpha_x^2}+1}}
\right)
=\frac{1}{8}\left(1\pm
\frac{\vert\left(\vert\alpha\vert^2-\vert\beta\vert^2\right)^2
-4\vert\alpha\beta\vert^2\sin^2\Delta\theta\vert}
{\left(\vert\alpha\vert^2-\vert\beta\vert^2\right)^2
+4\vert\alpha\beta\vert^2\sin^2\Delta\theta}
\right)
\end{eqnarray}
From equation~\eqref{eq:X^2_solutions} and using $\vert{TR}\vert^2=\vert{T}\vert^2-\vert{T}\vert^4=x^2$ allows us to compute the optimum transmission $\vert{T}\vert^2_{opt}$. We discard the solutions that yield a minimum Fisher information. The transmission coefficient corresponding to a maximum Fisher information is
\begin{equation}
\vert{T}\vert^2_{opt}
=\frac{1}{2}+\frac{\mathrm{sign}(\vert\beta\vert^2-\vert\alpha\vert^2)\vert\alpha\beta\vert \sin\Delta\theta}
{\sqrt{\left(\vert\alpha\vert^2-\vert\beta\vert^2\right)^2
+4\vert\alpha\beta\vert^2\sin^2\Delta\theta}
}
=\frac{1}{2}+
\frac{\mathrm{sign}(\varpi^2-1)\varpi\sin\Delta\theta}
{\sqrt{\left(1-\varpi^2\right)^2
+4\varpi^2\sin^2\Delta\theta}
}
\end{equation}

% ---------------------------------------------------------
% --------- DUAL COHERENT -- DIFFERENCE DETECTION ---------
% ---------------------------------------------------------
\section{Output observables for a difference intensity detection setup}
\label{sec:app:dual_coh_diff_det}
We close the setup from Fig.~\ref{fig:MZi_Fisher_two_phases} with a second beam splitter so that it becomes a MZI (see e. g. Fig.~2 from \citep{API18}).  We consider the input beam splitter having a transmission coefficient $T$ and the second one ($BS_2$) balanced. The output photo-detectors are assumed ideal and the difference photo-current is given by
\begin{eqnarray}
\label{eq:app:Nd_dual_coh_diff_det}
\hat{N}_d\left(\varphi\right)
=-2iTR^*\cos\varphi\hat{n}_0+2iTR^*\cos\varphi\hat{n}_1
+i\left(\vert{T}\vert^2e^{-i\varphi}
-\vert{R}\vert^2e^{i\varphi}
\right)\hat{a}_0^\dagger\hat{a}_1
+i\left(
-\vert{T}\vert^2e^{i\varphi}
+\vert{R}\vert^2e^{-i\varphi}
\right)\hat{a}_1^\dagger\hat{a}_0
\quad
\end{eqnarray}
The phase sensitivity \cite{Gar17,GerryKnight,API18,Dem15} using this setup is defined as
\begin{equation}
\label{eq:app:Delta_phi_dual_coh_diff_det}
\Delta\varphi_{\mathrm{diff}}
=\frac{\Delta\langle{N_d}\rangle}{\big\vert\frac{\partial\langle{N_d}\rangle}{\partial\varphi}\big\vert}
%=\frac{1}{\vert\alpha\vert}\times\frac{\sqrt{1+\varpi^2}}
%{2\big\vert\vert{TR}\vert\sin\varphi\left(1-\varpi^2\right)
%-2\varpi\cos\Delta\theta\cos\varphi
%+2\left(1-\vert{T}\vert^2\right)\varpi\sin\Delta\theta\sin\varphi\big\vert}
\end{equation}
%\end{widetext}
where $\Delta\langle{N_d}\rangle=\sqrt{\langle{\hat{N}_d^2}\rangle-\langle{\hat{N}_d}\rangle^2}$ and $\varphi$ is the total internal phase in the interferometer. A long but straightforward calculation yields the variance of the observable $\hat{N}_d$,
\begin{equation}
\label{eq:app:N_d_variance}
\Delta^2\langle{N_d}\rangle=\vert\alpha\vert^2\left(1+\varpi^2\right)
\end{equation} and from equation \eqref{eq:app:Nd_dual_coh_diff_det} we immediately have
\begin{eqnarray}
\label{eq:app:del_N_d_del_varphi}
\bigg\vert\frac{\partial\langle{N_d}\rangle}{\partial\varphi}\bigg\vert
=\vert\alpha\vert^2\big\vert2\vert{TR}\vert\sin\varphi\left(1-\varpi^2\right)
+2\varpi\cos\Delta\theta\cos\varphi
%\nonumber\\
+2\left(1-2\vert{T}\vert^2\right)\varpi\sin\Delta\theta\sin\varphi\big\vert
\qquad
\end{eqnarray}
Grouping together the terms depending on $\sin\varphi$ and making the notation
\begin{equation}
\label{eq:app:Notation_C_d}
C_d=\vert{TR}\vert\sin\varphi\left(1-\varpi^2\right)
+\left(1-2\vert{T}\vert^2\right)\varpi\sin\Delta\theta
\end{equation}
we arrive at the simple expression
\begin{eqnarray}
\label{eq:app:del_N_d_del_varphi2}
\bigg\vert\frac{\partial\langle{N_d}\rangle}{\partial\varphi}\bigg\vert
=2\vert\alpha\vert^2\vert C_d\sin\varphi
+\varpi\cos\Delta\theta\cos\varphi
\vert
\qquad
\end{eqnarray}

Combining the results from equations \eqref{eq:app:N_d_variance} and \eqref{eq:app:del_N_d_del_varphi2} with equation \eqref{eq:app:Delta_phi_dual_coh_diff_det} yields the result from equation \eqref{eq:Delta_phi_dual_coh_diff_det}.

Starting from equation \eqref{eq:Delta_phi_dual_coh_diff_det}, a short calculation shows that the optimum internal phase shift is given by
\begin{equation}
\label{eq:app:theta_opt_dial_coh_nonbal}
\varphi_\mathrm{opt}=\arctan\left(\frac{\left(1-2\vert{T}\vert^2\right)\varpi\sin\Delta\theta+\vert{TR}\vert(1-\varpi^2)}{\varpi\cos\Delta\theta}\right)+2k\pi
\end{equation}

% ---------------------------------------------------------
% --------- FISHER --- SQUEEZED plus COHERENT -------------
% ---------------------------------------------------------
\section{Fisher information calculation for the coherent plus squeezed vacuum input}
\label{sec:app:Fisher_calcualtion_ocherent_squeezed_vacuum}
For the sum-sum Fisher matrix element $\mathfrak{F}_{ss}$ we use the definition from equation \eqref{eq:Fisher_Matrix_elements_DEF} and have the first,
\begin{equation}
\langle\partial_s\psi\vert\partial_s\psi\rangle
=\frac{1}{4}\left(\vert\alpha\vert^4+\vert\alpha\vert^2+2\vert\alpha\vert^2\sinh^2r+2\sinh^4r+\sinh^2r\cosh^2r+\sinh^2r
\right)
\end{equation}
and respectively, second term $\langle\psi\vert\partial_s\psi\rangle
=-{i}/{2}\left(\vert\alpha\vert^2+\sinh^2r\right)$.
We obtain 
\begin{equation}
\label{eq:F_ss_sqz_vac_coh_input0}
\mathfrak{F}_{ss}
=\vert\alpha\vert^2+\frac{\sinh^22r}{2}
\end{equation}
For the first term of $\mathfrak{F}_{dd}$
we have
\begin{eqnarray}
\langle\partial_d\psi\vert\partial_d\psi\rangle
=\frac{1}{4}\left(2\sinh^4r+\sinh^2r\cosh^2r
-2\vert\alpha\vert^2\sinh^2r+\vert\alpha\vert^4
\right)
%\nonumber\\
% --------- SECOND LINE
-\vert{TR}\vert^2\left(
2\sinh^4r+\sinh^2r\cosh^2r+\vert\alpha\vert^4
\right)
\nonumber\\
% --------- THIRD LINE
+\frac{1}{4}\sinh^2r+\frac{1}{4}\vert\alpha\vert^2
+\vert{TR}\vert^2\vert\alpha\vert^2\sinh2r\cos\Delta\theta
+4\vert{TR}\vert^2\vert\alpha\vert^2\sinh^2r
\end{eqnarray}
while the second one is $\langle\psi\vert\partial_d\psi\rangle=-{i}/{2}\left(\vert{T}\vert^2-\vert{R}\vert^2\right)
\left(\sinh^2r-\vert\alpha\vert^2\right)
$, yielding the Fisher matrix element
\begin{eqnarray}
\label{eq:F_dd_sqz_vac_coh_input}
\mathfrak{F}_{dd}=\left(\vert{T}\vert^2-\vert{R}\vert^2\right)^2
\left(\vert\alpha\vert^2+\frac{\sinh^22r}{2}\right)
%\nonumber\\
% --------- SECOND LINE
+4\vert{TR}\vert^2\left(\vert\alpha\vert^2\left(\sinh2r\cos\Delta\theta+\cosh2r\right)+\sinh^2r\right)
\end{eqnarray}
The sum-difference and difference-sum Fisher matrix elements are equal and give
\begin{eqnarray}
\mathfrak{F}_{sd}=\mathfrak{F}_{ds}
=\left(\vert{T}\vert^2-\vert{R}\vert^2\right)
\left(\frac{\sinh^22r}{2}-\vert\alpha\vert^2
\right)
\end{eqnarray}
The determinant of the Fisher matrix is
$D_{\mathfrak{F}}=\mathfrak{F}_{ss}\mathfrak{F}_{dd}-\mathfrak{F}_{ds}\mathfrak{F}_{sd}$. The Fisher information for the difference-difference sensitivity measurement is $\mathcal{F}=D_{\mathfrak{F}}/\mathfrak{F}_{ss}$, yielding
\begin{eqnarray}
\label{eq:Fisher_sqz_vac_coh_input}
\mathcal{F}=\frac{2\left(\vert{T}\vert^2-\vert{R}\vert^2\right)^2\sinh^22r\vert\alpha\vert^2}
{\vert\alpha\vert^2+\frac{\sinh^22r}{2}}
%\nonumber\\
% --------- SECOND LINE
+4\vert{TR}\vert^2\left(\vert\alpha\vert^2\left(\sinh2r\cos\Delta\theta+\cosh2r\right)+\sinh^2r\right)
\end{eqnarray}
and using the identity $\left(\vert{T}\vert^2-\vert{R}\vert^2\right)^2=1-4\vert{TR}\vert^2$ takes us to the expression given in equation \eqref{eq:Fisher_coh_squeezed_vacuum_general}.

% ------------------------------------------
% ------------- FISHER --- SQUEEZED plus COHERENT ---------
% ---------------------------------------------------------
 \section{Fisher information calculation for the squeezed-coherent plus squeezed vacuum input}
\label{sec:app:Fisher_calcualtion_coh_sqz_plus_sqz_vac}
Following the definition of Fisher matrix elements \eqref{eq:Fisher_Matrix_elements_DEF} and the input state from equation \eqref{eq:psi_in_coh_squeezed_times_squeezed_vac} we obtain 
\begin{eqnarray}
\label{eq:app:F_ss_coh_sqz_vac_SQZ_VAC}
\mathfrak{F}_{ss}=\frac{\sinh^2{2r}}{2}
+\frac{\sinh^2{2z}}{2}
+\vert\alpha\vert^2\left(\cosh{2z}
-\sinh{2z}\cos{\Delta\phi}\right)
\end{eqnarray}
where we denoted the input phase mis-match between the coherent source and the squeezing in the same arm ${\Delta\phi=2\theta_\alpha-\phi}$.
In a similar manner we calculate the difference-difference Fisher matrix element
\begin{eqnarray}
\label{eq:F_dd_coh_sqz_vac_SQZ_VAC}
\mathfrak{F}_{dd}
=\left(\vert{T}\vert^2-\vert{R}\vert^2\right)^2
\left(\frac{\sinh^2{2r}}{2}
+\frac{\sinh^2{2z}}{2}
+\vert\alpha\vert^2\left(\cosh{2z}
-\sinh{2z}\cos\Delta\phi\right)
\right)
\nonumber\\
+4\vert{TR}\vert^2
\left(\vert\alpha\vert^2\left(\cosh{2r}
+\sinh{2r}\cos\Delta\theta
\right)
+\sinh^2{r}+\sinh^2{z}\right)
\nonumber\\
+8\vert{TR}\vert^2\sinh{r}\sinh{z}
\left(\sinh{r}\sinh{z}
-\cosh{r}\cosh{z}\cos(\phi-\theta)
\right)
\end{eqnarray}
and we remind the notation ${\Delta\theta=2\theta_\alpha-\theta}$ representing the input phase mis-match between the coherent source and the squeezed vacuum (see Section \ref{sec:Fisher_coh_sqz_vacuum}).
Since $\mathfrak{F}_{sd}=\mathfrak{F}_{ds}$ we compute only one of them. We find 
\begin{eqnarray}
\label{eq:F_sd_coh_sqz_vac_SQZ_VAC}
\mathfrak{F}_{sd}
=\left(\vert{T}\vert^2-\vert{R}\vert^2\right)
\left(\frac{\sinh^2{2r}}{2}
-\frac{\sinh^2{2z}}{2}
-\vert\alpha\vert^2\left(\cosh{2z}-\sinh{2z}\cos{\Delta\phi}\right)
\right)
\end{eqnarray}
We use now the definition from equation \eqref{eq:F_is_Fdd_minus_F_ds2_over_Fss} and obtain the Fisher information for the squeezed-coherent plus squeezed vacuum input,
\begin{eqnarray}
\label{eq:F_final_coh_sqz_vac_sqz_vac}
\mathcal{F}=4\vert{TR}\vert^2
\left(
\vert\alpha\vert^2\left(\cosh{2r}
+\sinh{2r}\cos\Delta\theta\right)
+\sinh^2{r}+\sinh^2{z}\right)
\nonumber\\
+8\vert{TR}\vert^2\sinh{r}\sinh{z}
\left(\sinh{r}\sinh{z}
-\cosh{r}\cosh{z}\cos(\phi-\theta)
\right)
\nonumber\\
+\left(\vert{T}\vert^2-\vert{R}\vert^2\right)^2
\frac{\sinh^2{2r}\left(\sinh^2{2z}+2\vert\alpha\vert^2\left(
\cosh2z-\sinh2z\cos\Delta\phi\right)\right)}
{\frac{\sinh^2{2r}}{2}+\frac{\sinh^22z}{2}+\vert\alpha\vert^2\left(
\cosh2z-\sinh2z\cos\Delta\phi\right)}
\end{eqnarray}
For the balanced case one obtains the result from equation \eqref{eq:F_final_coh_sqz_vac_sqz_vac_BALANCED}.

If we want to impose on $\mathcal{F}$ no $T$-dependence, we need to satisfy the condition $\kappa=0$ where we define
\begin{eqnarray}
\label{eq:app:kappa_zero_coh_sqz_vac_sqz_vac}
\kappa
=\vert\alpha\vert^2
\left(
\left(\cosh{2r}+\sinh{2r}\cos\Delta\theta\right)
+\sinh^2{r}+\sinh^2{z}\right)
%\nonumber\\
+2\sinh{r}\sinh{z}
\left(\sinh{r}\sinh{z}
-\cosh{r}\cosh{z}\cos(\phi-\theta)\right)
\nonumber
\\
-\frac{\sinh^2{2r}\left(\sinh^2{2z}+2\vert\alpha\vert^2\left(
\cosh2z-\sinh2z\cos\Delta\phi\right)\right)}
{\frac{\sinh^2{2r}}{2}+\frac{\sinh^22z}{2}+\vert\alpha\vert^2\left(
\cosh2z-\sinh2z\cos\Delta\phi\right)}
\qquad
\end{eqnarray}

\end{widetext}

% #########################################################
% #############    B I B L I O G R A P H Y    #############
% #########################################################
%
% BibTeX users please use
% \bibliographystyle{}
% \bibliography{}
%
% APS style
%\bibliographystyle{apsrev4-1}

% now include the BIB file
\bibliography{MZI_phase_sensitivity_bibtex}

\begin{thebibliography}{35}%
\makeatletter
\providecommand \@ifxundefined [1]{%
 \@ifx{#1\undefined}
}%
\providecommand \@ifnum [1]{%
 \ifnum #1\expandafter \@firstoftwo
 \else \expandafter \@secondoftwo
 \fi
}%
\providecommand \@ifx [1]{%
 \ifx #1\expandafter \@firstoftwo
 \else \expandafter \@secondoftwo
 \fi
}%
\providecommand \natexlab [1]{#1}%
\providecommand \enquote  [1]{``#1''}%
\providecommand \bibnamefont  [1]{#1}%
\providecommand \bibfnamefont [1]{#1}%
\providecommand \citenamefont [1]{#1}%
\providecommand \href@noop [0]{\@secondoftwo}%
\providecommand \href [0]{\begingroup \@sanitize@url \@href}%
\providecommand \@href[1]{\@@startlink{#1}\@@href}%
\providecommand \@@href[1]{\endgroup#1\@@endlink}%
\providecommand \@sanitize@url [0]{\catcode `\\12\catcode `\$12\catcode
  `\&12\catcode `\#12\catcode `\^12\catcode `\_12\catcode `\%12\relax}%
\providecommand \@@startlink[1]{}%
\providecommand \@@endlink[0]{}%
\providecommand \url  [0]{\begingroup\@sanitize@url \@url }%
\providecommand \@url [1]{\endgroup\@href {#1}{\urlprefix }}%
\providecommand \urlprefix  [0]{URL }%
\providecommand \Eprint [0]{\href }%
\providecommand \doibase [0]{http://dx.doi.org/}%
\providecommand \selectlanguage [0]{\@gobble}%
\providecommand \bibinfo  [0]{\@secondoftwo}%
\providecommand \bibfield  [0]{\@secondoftwo}%
\providecommand \translation [1]{[#1]}%
\providecommand \BibitemOpen [0]{}%
\providecommand \bibitemStop [0]{}%
\providecommand \bibitemNoStop [0]{.\EOS\space}%
\providecommand \EOS [0]{\spacefactor3000\relax}%
\providecommand \BibitemShut  [1]{\csname bibitem#1\endcsname}%
\let\auto@bib@innerbib\@empty
%</preamble>
\bibitem [{\citenamefont {Paris}(2009)}]{Par09}%
  \BibitemOpen
  \bibfield  {author} {\bibinfo {author} {\bibfnamefont {M.~G.~A.}\
  \bibnamefont {Paris}},\ }\href {\doibase 10.1142/S0219749909004839}
  {\bibfield  {journal} {\bibinfo  {journal} {Int. J. Quant. Info.}\ }\textbf
  {\bibinfo {volume} {07}},\ \bibinfo {pages} {125} (\bibinfo {year}
  {2009})}\BibitemShut {NoStop}%
\bibitem [{\citenamefont {Giovannetti}\ \emph {et~al.}(2011)\citenamefont
  {Giovannetti}, \citenamefont {Lloyd},\ and\ \citenamefont {Maccone}}]{Gio11}%
  \BibitemOpen
  \bibfield  {author} {\bibinfo {author} {\bibfnamefont {V.}~\bibnamefont
  {Giovannetti}}, \bibinfo {author} {\bibfnamefont {S.}~\bibnamefont {Lloyd}},
  \ and\ \bibinfo {author} {\bibfnamefont {L.}~\bibnamefont {Maccone}},\ }\href
  {\doibase 10.1038/nphoton.2011.35} {\bibfield  {journal} {\bibinfo  {journal}
  {Nature Photonics}\ }\textbf {\bibinfo {volume} {5}},\ \bibinfo {pages} {222}
  (\bibinfo {year} {2011})}\BibitemShut {NoStop}%
\bibitem [{\citenamefont {Jarzyna}\ and\ \citenamefont
  {Demkowicz-Dobrza\'{n}ski}(2015)}]{Jar15}%
  \BibitemOpen
  \bibfield  {author} {\bibinfo {author} {\bibfnamefont {M.}~\bibnamefont
  {Jarzyna}}\ and\ \bibinfo {author} {\bibfnamefont {R.}~\bibnamefont
  {Demkowicz-Dobrza\'{n}ski}},\ }\href {\doibase 10.1088/1367-2630/17/1/013010}
  {\bibfield  {journal} {\bibinfo  {journal} {New Journal of Physics}\ }\textbf
  {\bibinfo {volume} {17}},\ \bibinfo {pages} {013010} (\bibinfo {year}
  {2015})}\BibitemShut {NoStop}%
\bibitem [{\citenamefont {Ionicioiu}(2015)}]{Ion15}%
  \BibitemOpen
  \bibfield  {author} {\bibinfo {author} {\bibfnamefont {R.}~\bibnamefont
  {Ionicioiu}},\ }\href@noop {} {\bibfield  {journal} {\bibinfo  {journal}
  {Rom. Rep. Phys.}\ }\textbf {\bibinfo {volume} {67}},\ \bibinfo {pages}
  {1300} (\bibinfo {year} {2015})}\BibitemShut {NoStop}%
\bibitem [{\citenamefont {{The LIGO Scientific Collaboration}}(2013)}]{LIGO13}%
  \BibitemOpen
  \bibfield  {author} {\bibinfo {author} {\bibnamefont {{The LIGO Scientific
  Collaboration}}},\ }\href {\doibase 10.1038/NPHOTON.2013.177} {\bibfield
  {journal} {\bibinfo  {journal} {Nature Photonics}\ }\textbf {\bibinfo
  {volume} {7}},\ \bibinfo {pages} {616} (\bibinfo {year} {2013})}\BibitemShut
  {NoStop}%
\bibitem [{\citenamefont {Demkowicz-Dobrza\ifmmode~\acute{n}\else
  \'{n}\fi{}ski}\ \emph {et~al.}(2013)\citenamefont
  {Demkowicz-Dobrza\ifmmode~\acute{n}\else \'{n}\fi{}ski}, \citenamefont
  {Banaszek},\ and\ \citenamefont {Schnabel}}]{Dem13}%
  \BibitemOpen
  \bibfield  {author} {\bibinfo {author} {\bibfnamefont {R.}~\bibnamefont
  {Demkowicz-Dobrza\ifmmode~\acute{n}\else \'{n}\fi{}ski}}, \bibinfo {author}
  {\bibfnamefont {K.}~\bibnamefont {Banaszek}}, \ and\ \bibinfo {author}
  {\bibfnamefont {R.}~\bibnamefont {Schnabel}},\ }\href {\doibase
  10.1103/PhysRevA.88.041802} {\bibfield  {journal} {\bibinfo  {journal} {Phys.
  Rev. A}\ }\textbf {\bibinfo {volume} {88}},\ \bibinfo {pages} {041802(R)}
  (\bibinfo {year} {2013})}\BibitemShut {NoStop}%
\bibitem [{\citenamefont {Schnabel}(2017)}]{Sch17}%
  \BibitemOpen
  \bibfield  {author} {\bibinfo {author} {\bibfnamefont {R.}~\bibnamefont
  {Schnabel}},\ }\href {\doibase 10.1016/j.physrep.2017.04.001} {\bibfield
  {journal} {\bibinfo  {journal} {Physics Reports}\ }\textbf {\bibinfo {volume}
  {684}},\ \bibinfo {pages} {1 } (\bibinfo {year} {2017})}\BibitemShut
  {NoStop}%
\bibitem [{\citenamefont {Ataman}(2018)}]{Ata18b}%
  \BibitemOpen
  \bibfield  {author} {\bibinfo {author} {\bibfnamefont {S.}~\bibnamefont
  {Ataman}},\ }\href {\doibase 10.1103/PhysRevA.97.063811} {\bibfield
  {journal} {\bibinfo  {journal} {Phys. Rev. A}\ }\textbf {\bibinfo {volume}
  {97}},\ \bibinfo {pages} {063811} (\bibinfo {year} {2018})}\BibitemShut
  {NoStop}%
\bibitem [{\citenamefont {Braunstein}\ and\ \citenamefont
  {Caves}(1994)}]{Bra94}%
  \BibitemOpen
  \bibfield  {author} {\bibinfo {author} {\bibfnamefont {S.~L.}\ \bibnamefont
  {Braunstein}}\ and\ \bibinfo {author} {\bibfnamefont {C.~M.}\ \bibnamefont
  {Caves}},\ }\href {\doibase 10.1103/PhysRevLett.72.3439} {\bibfield
  {journal} {\bibinfo  {journal} {Phys. Rev. Lett.}\ }\textbf {\bibinfo
  {volume} {72}},\ \bibinfo {pages} {3439} (\bibinfo {year}
  {1994})}\BibitemShut {NoStop}%
\bibitem [{\citenamefont {Demkowicz-Dobrza\'{n}ski}\ \emph
  {et~al.}(2015)\citenamefont {Demkowicz-Dobrza\'{n}ski}, \citenamefont
  {Jarzyna},\ and\ \citenamefont {Ko\l{}ody\'{n}ski}}]{Dem15}%
  \BibitemOpen
  \bibfield  {author} {\bibinfo {author} {\bibfnamefont {R.}~\bibnamefont
  {Demkowicz-Dobrza\'{n}ski}}, \bibinfo {author} {\bibfnamefont
  {M.}~\bibnamefont {Jarzyna}}, \ and\ \bibinfo {author} {\bibfnamefont
  {J.}~\bibnamefont {Ko\l{}ody\'{n}ski}},\ }\href {\doibase
  10.1016/bs.po.2015.02.003} {\bibfield  {journal} {\bibinfo  {journal}
  {Progress in Optics}\ }\textbf {\bibinfo {volume} {60}},\ \bibinfo {pages}
  {345 } (\bibinfo {year} {2015})}\BibitemShut {NoStop}%
\bibitem [{\citenamefont {Caves}(1981)}]{Cav81}%
  \BibitemOpen
  \bibfield  {author} {\bibinfo {author} {\bibfnamefont {C.~M.}\ \bibnamefont
  {Caves}},\ }\href {\doibase 10.1103/PhysRevD.23.1693} {\bibfield  {journal}
  {\bibinfo  {journal} {Phys. Rev. D}\ }\textbf {\bibinfo {volume} {23}},\
  \bibinfo {pages} {1693} (\bibinfo {year} {1981})}\BibitemShut {NoStop}%
\bibitem [{\citenamefont {Giovannetti}\ and\ \citenamefont
  {Maccone}(2012)}]{Gio12}%
  \BibitemOpen
  \bibfield  {author} {\bibinfo {author} {\bibfnamefont {V.}~\bibnamefont
  {Giovannetti}}\ and\ \bibinfo {author} {\bibfnamefont {L.}~\bibnamefont
  {Maccone}},\ }\href {\doibase 10.1103/PhysRevLett.108.210404} {\bibfield
  {journal} {\bibinfo  {journal} {Phys. Rev. Lett.}\ }\textbf {\bibinfo
  {volume} {108}},\ \bibinfo {pages} {210404} (\bibinfo {year}
  {2012})}\BibitemShut {NoStop}%
\bibitem [{\citenamefont {Holland}\ and\ \citenamefont
  {Burnett}(1993)}]{Hol93}%
  \BibitemOpen
  \bibfield  {author} {\bibinfo {author} {\bibfnamefont {M.~J.}\ \bibnamefont
  {Holland}}\ and\ \bibinfo {author} {\bibfnamefont {K.}~\bibnamefont
  {Burnett}},\ }\href {\doibase 10.1103/PhysRevLett.71.1355} {\bibfield
  {journal} {\bibinfo  {journal} {Phys. Rev. Lett.}\ }\textbf {\bibinfo
  {volume} {71}},\ \bibinfo {pages} {1355} (\bibinfo {year}
  {1993})}\BibitemShut {NoStop}%
\bibitem [{\citenamefont {Boto}\ \emph {et~al.}(2000)\citenamefont {Boto},
  \citenamefont {Kok}, \citenamefont {Abrams}, \citenamefont {Braunstein},
  \citenamefont {Williams},\ and\ \citenamefont {Dowling}}]{Bot00}%
  \BibitemOpen
  \bibfield  {author} {\bibinfo {author} {\bibfnamefont {A.~N.}\ \bibnamefont
  {Boto}}, \bibinfo {author} {\bibfnamefont {P.}~\bibnamefont {Kok}}, \bibinfo
  {author} {\bibfnamefont {D.~S.}\ \bibnamefont {Abrams}}, \bibinfo {author}
  {\bibfnamefont {S.~L.}\ \bibnamefont {Braunstein}}, \bibinfo {author}
  {\bibfnamefont {C.~P.}\ \bibnamefont {Williams}}, \ and\ \bibinfo {author}
  {\bibfnamefont {J.~P.}\ \bibnamefont {Dowling}},\ }\href {\doibase
  10.1103/PhysRevLett.85.2733} {\bibfield  {journal} {\bibinfo  {journal}
  {Phys. Rev. Lett.}\ }\textbf {\bibinfo {volume} {85}},\ \bibinfo {pages}
  {2733} (\bibinfo {year} {2000})}\BibitemShut {NoStop}%
\bibitem [{\citenamefont {Campos}\ \emph {et~al.}(2003)\citenamefont {Campos},
  \citenamefont {Gerry},\ and\ \citenamefont {Benmoussa}}]{Cam03}%
  \BibitemOpen
  \bibfield  {author} {\bibinfo {author} {\bibfnamefont {R.~A.}\ \bibnamefont
  {Campos}}, \bibinfo {author} {\bibfnamefont {C.~C.}\ \bibnamefont {Gerry}}, \
  and\ \bibinfo {author} {\bibfnamefont {A.}~\bibnamefont {Benmoussa}},\ }\href
  {\doibase 10.1103/PhysRevA.68.023810} {\bibfield  {journal} {\bibinfo
  {journal} {Phys. Rev. A}\ }\textbf {\bibinfo {volume} {68}},\ \bibinfo
  {pages} {023810} (\bibinfo {year} {2003})}\BibitemShut {NoStop}%
\bibitem [{\citenamefont {Dorner}\ \emph {et~al.}(2009)\citenamefont {Dorner},
  \citenamefont {Demkowicz-Dobrzanski}, \citenamefont {Smith}, \citenamefont
  {Lundeen}, \citenamefont {Wasilewski}, \citenamefont {Banaszek},\ and\
  \citenamefont {Walmsley}}]{Dor09}%
  \BibitemOpen
  \bibfield  {author} {\bibinfo {author} {\bibfnamefont {U.}~\bibnamefont
  {Dorner}}, \bibinfo {author} {\bibfnamefont {R.}~\bibnamefont
  {Demkowicz-Dobrzanski}}, \bibinfo {author} {\bibfnamefont {B.~J.}\
  \bibnamefont {Smith}}, \bibinfo {author} {\bibfnamefont {J.~S.}\ \bibnamefont
  {Lundeen}}, \bibinfo {author} {\bibfnamefont {W.}~\bibnamefont {Wasilewski}},
  \bibinfo {author} {\bibfnamefont {K.}~\bibnamefont {Banaszek}}, \ and\
  \bibinfo {author} {\bibfnamefont {I.~A.}\ \bibnamefont {Walmsley}},\ }\href
  {\doibase 10.1103/PhysRevLett.102.040403} {\bibfield  {journal} {\bibinfo
  {journal} {Phys. Rev. Lett.}\ }\textbf {\bibinfo {volume} {102}},\ \bibinfo
  {pages} {040403} (\bibinfo {year} {2009})}\BibitemShut {NoStop}%
\bibitem [{\citenamefont {Demkowicz-Dobrza\'{n}ski}\ \emph
  {et~al.}(2012)\citenamefont {Demkowicz-Dobrza\'{n}ski}, \citenamefont
  {Ko\l{}ody\'{n}ski},\ and\ \citenamefont {Gu\c{t}\u{a}}}]{Dem12}%
  \BibitemOpen
  \bibfield  {author} {\bibinfo {author} {\bibfnamefont {R.}~\bibnamefont
  {Demkowicz-Dobrza\'{n}ski}}, \bibinfo {author} {\bibfnamefont
  {J.}~\bibnamefont {Ko\l{}ody\'{n}ski}}, \ and\ \bibinfo {author}
  {\bibfnamefont {M.}~\bibnamefont {Gu\c{t}\u{a}}},\ }\href {\doibase
  10.1038/ncomms2067} {\bibfield  {journal} {\bibinfo  {journal} {Nature
  Communications}\ }\textbf {\bibinfo {volume} {3}},\ \bibinfo {pages} {1063}
  (\bibinfo {year} {2012})}\BibitemShut {NoStop}%
\bibitem [{\citenamefont {Yuen}(1976)}]{Yue76}%
  \BibitemOpen
  \bibfield  {author} {\bibinfo {author} {\bibfnamefont {H.~P.}\ \bibnamefont
  {Yuen}},\ }\href {\doibase 10.1103/PhysRevA.13.2226} {\bibfield  {journal}
  {\bibinfo  {journal} {Phys. Rev. A}\ }\textbf {\bibinfo {volume} {13}},\
  \bibinfo {pages} {2226} (\bibinfo {year} {1976})}\BibitemShut {NoStop}%
\bibitem [{\citenamefont {Yurke}(1985)}]{Yur85}%
  \BibitemOpen
  \bibfield  {author} {\bibinfo {author} {\bibfnamefont {B.}~\bibnamefont
  {Yurke}},\ }\href {\doibase 10.1103/PhysRevA.32.300} {\bibfield  {journal}
  {\bibinfo  {journal} {Phys. Rev. A}\ }\textbf {\bibinfo {volume} {32}},\
  \bibinfo {pages} {300} (\bibinfo {year} {1985})}\BibitemShut {NoStop}%
\bibitem [{\citenamefont {Xiao}\ \emph {et~al.}(1987)\citenamefont {Xiao},
  \citenamefont {Wu},\ and\ \citenamefont {Kimble}}]{Xia87}%
  \BibitemOpen
  \bibfield  {author} {\bibinfo {author} {\bibfnamefont {M.}~\bibnamefont
  {Xiao}}, \bibinfo {author} {\bibfnamefont {L.-A.}\ \bibnamefont {Wu}}, \ and\
  \bibinfo {author} {\bibfnamefont {H.~J.}\ \bibnamefont {Kimble}},\ }\href
  {\doibase 10.1103/PhysRevLett.59.278} {\bibfield  {journal} {\bibinfo
  {journal} {Phys. Rev. Lett.}\ }\textbf {\bibinfo {volume} {59}},\ \bibinfo
  {pages} {278} (\bibinfo {year} {1987})}\BibitemShut {NoStop}%
\bibitem [{\citenamefont {Gard}\ \emph {et~al.}(2017)\citenamefont {Gard},
  \citenamefont {You}, \citenamefont {Mishra}, \citenamefont {Singh},
  \citenamefont {Lee}, \citenamefont {Corbitt},\ and\ \citenamefont
  {Dowling}}]{Gar17}%
  \BibitemOpen
  \bibfield  {author} {\bibinfo {author} {\bibfnamefont {B.~T.}\ \bibnamefont
  {Gard}}, \bibinfo {author} {\bibfnamefont {C.}~\bibnamefont {You}}, \bibinfo
  {author} {\bibfnamefont {D.~K.}\ \bibnamefont {Mishra}}, \bibinfo {author}
  {\bibfnamefont {R.}~\bibnamefont {Singh}}, \bibinfo {author} {\bibfnamefont
  {H.}~\bibnamefont {Lee}}, \bibinfo {author} {\bibfnamefont {T.~R.}\
  \bibnamefont {Corbitt}}, \ and\ \bibinfo {author} {\bibfnamefont {J.~P.}\
  \bibnamefont {Dowling}},\ }\href {\doibase 10.1140/epjqt/s40507-017-0058-8}
  {\bibfield  {journal} {\bibinfo  {journal} {EPJ Quantum Technology}\ }\textbf
  {\bibinfo {volume} {4}},\ \bibinfo {pages} {4} (\bibinfo {year}
  {2017})}\BibitemShut {NoStop}%
\bibitem [{\citenamefont {Ataman}\ \emph {et~al.}(2018)\citenamefont {Ataman},
  \citenamefont {Preda},\ and\ \citenamefont {Ionicioiu}}]{API18}%
  \BibitemOpen
  \bibfield  {author} {\bibinfo {author} {\bibfnamefont {S.}~\bibnamefont
  {Ataman}}, \bibinfo {author} {\bibfnamefont {A.}~\bibnamefont {Preda}}, \
  and\ \bibinfo {author} {\bibfnamefont {R.}~\bibnamefont {Ionicioiu}},\ }\href
  {\doibase 10.1103/PhysRevA.98.043856} {\bibfield  {journal} {\bibinfo
  {journal} {Phys. Rev. A}\ }\textbf {\bibinfo {volume} {98}},\ \bibinfo
  {pages} {043856} (\bibinfo {year} {2018})}\BibitemShut {NoStop}%
\bibitem [{\citenamefont {Pezz\'e}\ \emph {et~al.}(2007)\citenamefont
  {Pezz\'e}, \citenamefont {Smerzi}, \citenamefont {Khoury}, \citenamefont
  {Hodelin},\ and\ \citenamefont {Bouwmeester}}]{Pez07}%
  \BibitemOpen
  \bibfield  {author} {\bibinfo {author} {\bibfnamefont {L.}~\bibnamefont
  {Pezz\'e}}, \bibinfo {author} {\bibfnamefont {A.}~\bibnamefont {Smerzi}},
  \bibinfo {author} {\bibfnamefont {G.}~\bibnamefont {Khoury}}, \bibinfo
  {author} {\bibfnamefont {J.~F.}\ \bibnamefont {Hodelin}}, \ and\ \bibinfo
  {author} {\bibfnamefont {D.}~\bibnamefont {Bouwmeester}},\ }\href {\doibase
  10.1103/PhysRevLett.99.223602} {\bibfield  {journal} {\bibinfo  {journal}
  {Phys. Rev. Lett.}\ }\textbf {\bibinfo {volume} {99}},\ \bibinfo {pages}
  {223602} (\bibinfo {year} {2007})}\BibitemShut {NoStop}%
\bibitem [{\citenamefont {Jarzyna}\ and\ \citenamefont
  {Demkowicz-Dobrza\ifmmode~\acute{n}\else \'{n}\fi{}ski}(2012)}]{Jar12}%
  \BibitemOpen
  \bibfield  {author} {\bibinfo {author} {\bibfnamefont {M.}~\bibnamefont
  {Jarzyna}}\ and\ \bibinfo {author} {\bibfnamefont {R.}~\bibnamefont
  {Demkowicz-Dobrza\ifmmode~\acute{n}\else \'{n}\fi{}ski}},\ }\href {\doibase
  10.1103/PhysRevA.85.011801} {\bibfield  {journal} {\bibinfo  {journal} {Phys.
  Rev. A}\ }\textbf {\bibinfo {volume} {85}},\ \bibinfo {pages} {011801(R)}
  (\bibinfo {year} {2012})}\BibitemShut {NoStop}%
\bibitem [{\citenamefont {Takeoka}\ \emph {et~al.}(2017)\citenamefont
  {Takeoka}, \citenamefont {Seshadreesan}, \citenamefont {You}, \citenamefont
  {Izumi},\ and\ \citenamefont {Dowling}}]{Tak17}%
  \BibitemOpen
  \bibfield  {author} {\bibinfo {author} {\bibfnamefont {M.}~\bibnamefont
  {Takeoka}}, \bibinfo {author} {\bibfnamefont {K.~P.}\ \bibnamefont
  {Seshadreesan}}, \bibinfo {author} {\bibfnamefont {C.}~\bibnamefont {You}},
  \bibinfo {author} {\bibfnamefont {S.}~\bibnamefont {Izumi}}, \ and\ \bibinfo
  {author} {\bibfnamefont {J.~P.}\ \bibnamefont {Dowling}},\ }\href {\doibase
  10.1103/PhysRevA.96.052118} {\bibfield  {journal} {\bibinfo  {journal} {Phys.
  Rev. A}\ }\textbf {\bibinfo {volume} {96}},\ \bibinfo {pages} {052118}
  (\bibinfo {year} {2017})}\BibitemShut {NoStop}%
\bibitem [{\citenamefont {Lang}\ and\ \citenamefont {Caves}(2013)}]{Lan13}%
  \BibitemOpen
  \bibfield  {author} {\bibinfo {author} {\bibfnamefont {M.~D.}\ \bibnamefont
  {Lang}}\ and\ \bibinfo {author} {\bibfnamefont {C.~M.}\ \bibnamefont
  {Caves}},\ }\href {\doibase 10.1103/PhysRevLett.111.173601} {\bibfield
  {journal} {\bibinfo  {journal} {Phys. Rev. Lett.}\ }\textbf {\bibinfo
  {volume} {111}},\ \bibinfo {pages} {173601} (\bibinfo {year}
  {2013})}\BibitemShut {NoStop}%
\bibitem [{\citenamefont {Lang}\ and\ \citenamefont {Caves}(2014)}]{Lan14}%
  \BibitemOpen
  \bibfield  {author} {\bibinfo {author} {\bibfnamefont {M.~D.}\ \bibnamefont
  {Lang}}\ and\ \bibinfo {author} {\bibfnamefont {C.~M.}\ \bibnamefont
  {Caves}},\ }\href {\doibase 10.1103/PhysRevA.90.025802} {\bibfield  {journal}
  {\bibinfo  {journal} {Phys. Rev. A}\ }\textbf {\bibinfo {volume} {90}},\
  \bibinfo {pages} {025802} (\bibinfo {year} {2014})}\BibitemShut {NoStop}%
\bibitem [{\citenamefont {Liu}\ \emph {et~al.}(2013)\citenamefont {Liu},
  \citenamefont {Jing},\ and\ \citenamefont {Wang}}]{Liu13}%
  \BibitemOpen
  \bibfield  {author} {\bibinfo {author} {\bibfnamefont {J.}~\bibnamefont
  {Liu}}, \bibinfo {author} {\bibfnamefont {X.}~\bibnamefont {Jing}}, \ and\
  \bibinfo {author} {\bibfnamefont {X.}~\bibnamefont {Wang}},\ }\href {\doibase
  10.1103/PhysRevA.88.042316} {\bibfield  {journal} {\bibinfo  {journal} {Phys.
  Rev. A}\ }\textbf {\bibinfo {volume} {88}},\ \bibinfo {pages} {042316}
  (\bibinfo {year} {2013})}\BibitemShut {NoStop}%
\bibitem [{\citenamefont {Pezz\`e}\ \emph {et~al.}(2015)\citenamefont
  {Pezz\`e}, \citenamefont {Hyllus},\ and\ \citenamefont {Smerzi}}]{Pez15}%
  \BibitemOpen
  \bibfield  {author} {\bibinfo {author} {\bibfnamefont {L.}~\bibnamefont
  {Pezz\`e}}, \bibinfo {author} {\bibfnamefont {P.}~\bibnamefont {Hyllus}}, \
  and\ \bibinfo {author} {\bibfnamefont {A.}~\bibnamefont {Smerzi}},\ }\href
  {\doibase 10.1103/PhysRevA.91.032103} {\bibfield  {journal} {\bibinfo
  {journal} {Phys. Rev. A}\ }\textbf {\bibinfo {volume} {91}},\ \bibinfo
  {pages} {032103} (\bibinfo {year} {2015})}\BibitemShut {NoStop}%
\bibitem [{\citenamefont {Shin}\ \emph {et~al.}(1999)\citenamefont {Shin},
  \citenamefont {Kim}, \citenamefont {Park}, \citenamefont {Kim},\ and\
  \citenamefont {Park}}]{Shi99}%
  \BibitemOpen
  \bibfield  {author} {\bibinfo {author} {\bibfnamefont {J.-T.}\ \bibnamefont
  {Shin}}, \bibinfo {author} {\bibfnamefont {H.-N.}\ \bibnamefont {Kim}},
  \bibinfo {author} {\bibfnamefont {G.-D.}\ \bibnamefont {Park}}, \bibinfo
  {author} {\bibfnamefont {T.-S.}\ \bibnamefont {Kim}}, \ and\ \bibinfo
  {author} {\bibfnamefont {D.-Y.}\ \bibnamefont {Park}},\ }\href
  {http://www.osapublishing.org/josk/abstract.cfm?URI=josk-3-1-1} {\bibfield
  {journal} {\bibinfo  {journal} {J. Opt. Soc. Korea}\ }\textbf {\bibinfo
  {volume} {3}},\ \bibinfo {pages} {1} (\bibinfo {year} {1999})}\BibitemShut
  {NoStop}%
\bibitem [{\citenamefont {Sparaciari}\ \emph {et~al.}(2015)\citenamefont
  {Sparaciari}, \citenamefont {Olivares},\ and\ \citenamefont {Paris}}]{Spa15}%
  \BibitemOpen
  \bibfield  {author} {\bibinfo {author} {\bibfnamefont {C.}~\bibnamefont
  {Sparaciari}}, \bibinfo {author} {\bibfnamefont {S.}~\bibnamefont
  {Olivares}}, \ and\ \bibinfo {author} {\bibfnamefont {M.~G.~A.}\ \bibnamefont
  {Paris}},\ }\href {\doibase 10.1364/JOSAB.32.001354} {\bibfield  {journal}
  {\bibinfo  {journal} {J. Opt. Soc. Am. B}\ }\textbf {\bibinfo {volume}
  {32}},\ \bibinfo {pages} {1354} (\bibinfo {year} {2015})}\BibitemShut
  {NoStop}%
\bibitem [{\citenamefont {Sparaciari}\ \emph {et~al.}(2016)\citenamefont
  {Sparaciari}, \citenamefont {Olivares},\ and\ \citenamefont {Paris}}]{Spa16}%
  \BibitemOpen
  \bibfield  {author} {\bibinfo {author} {\bibfnamefont {C.}~\bibnamefont
  {Sparaciari}}, \bibinfo {author} {\bibfnamefont {S.}~\bibnamefont
  {Olivares}}, \ and\ \bibinfo {author} {\bibfnamefont {M.~G.~A.}\ \bibnamefont
  {Paris}},\ }\href {\doibase 10.1103/PhysRevA.93.023810} {\bibfield  {journal}
  {\bibinfo  {journal} {Phys. Rev. A}\ }\textbf {\bibinfo {volume} {93}},\
  \bibinfo {pages} {023810} (\bibinfo {year} {2016})}\BibitemShut {NoStop}%
\bibitem [{\citenamefont {Paris}(1995)}]{Par95}%
  \BibitemOpen
  \bibfield  {author} {\bibinfo {author} {\bibfnamefont {M.~G.}\ \bibnamefont
  {Paris}},\ }\href {\doibase 10.1016/0375-9601(95)00235-U} {\bibfield
  {journal} {\bibinfo  {journal} {Physics Letters A}\ }\textbf {\bibinfo
  {volume} {201}},\ \bibinfo {pages} {132 } (\bibinfo {year}
  {1995})}\BibitemShut {NoStop}%
\bibitem [{\citenamefont {Gerry}\ and\ \citenamefont
  {Knight}(2005)}]{GerryKnight}%
  \BibitemOpen
  \bibfield  {author} {\bibinfo {author} {\bibfnamefont {C.}~\bibnamefont
  {Gerry}}\ and\ \bibinfo {author} {\bibfnamefont {P.}~\bibnamefont {Knight}},\
  }\href {\doibase 10.1017/CBO9780511791239} {\emph {\bibinfo {title}
  {Introductory Quantum Optics}}}\ (\bibinfo  {publisher} {Cambridge University
  Press},\ \bibinfo {year} {2005})\BibitemShut {NoStop}%
\bibitem [{\citenamefont {Pezz\'e}\ and\ \citenamefont {Smerzi}(2008)}]{Pez08}%
  \BibitemOpen
  \bibfield  {author} {\bibinfo {author} {\bibfnamefont {L.}~\bibnamefont
  {Pezz\'e}}\ and\ \bibinfo {author} {\bibfnamefont {A.}~\bibnamefont
  {Smerzi}},\ }\href {\doibase 10.1103/PhysRevLett.100.073601} {\bibfield
  {journal} {\bibinfo  {journal} {Phys. Rev. Lett.}\ }\textbf {\bibinfo
  {volume} {100}},\ \bibinfo {pages} {073601} (\bibinfo {year}
  {2008})}\BibitemShut {NoStop}%
\end{thebibliography}%

\end{document}